\documentclass[numsec,webpdf,contemporary,large]{oup-authoring-template}%

\onecolumn 

\graphicspath{{Fig/}}

\newcommand{\Multinomial}{\operatorname{Multinomial}}
\newcommand{\Binom}{\operatorname{Binom}}

\usepackage{dsfont}
\DeclareMathAlphabet{\mathcal}{OMS}{cmsy}{m}{n}
\usepackage{hyperref}

\theoremstyle{thmstyleone}%
\theoremstyle{thmstyletwo}%
\theoremstyle{thmstylethree}%

\begin{document}

\journaltitle{Arxiv}
\DOI{Hi1}
\copyrightyear{2024}
\pubyear{1}
\access{2}
\appnotes{3}

\firstpage{1}


\title[A Bayesian Approach for Earthquake Impact Modelling]{A Bayesian Approach for Earthquake Impact Modelling}

\author[1,$\ast$]{Max Anderson Loake\ORCID{0000-0002-0581-1617}}
\author[2]{Hamish Patten\ORCID{0000-0001-5600-2097}}
\author[1]{David Steinsaltz\ORCID{0000-0003-3044-5433}}

\authormark{Anderson Loake, Patten and Steinsaltz}

\address[1]{\orgdiv{Department of Statistics}, \orgname{University of Oxford}, \orgaddress{\street{St Giles'}, \postcode{OX1 3LB}, \state{Oxfordshire}, \country{United Kingdom}}} 

\address[2]{\orgdiv{Statistics Division}, \orgname{United Nations Food and Agriculture Organization}, \orgaddress{\country{Italy}}}

 \corresp[$\ast$]{Corresponding author. \href{email:email-id.com}{max.andersonloake@keble.ox.ac.uk}}




\abstract{
Immediately following a disaster event, such as an earthquake, estimates of the damage extent play a key role in informing the coordination of response and recovery efforts. We develop a novel impact estimation tool that leverages a generalised Bayesian approach to generate earthquake impact estimates across three impact types: mortality, population displacement, and building damage. Inference is performed within a likelihood-free framework, and a scoring-rule-based posterior avoids information loss from non-sufficient summary statistics. We propose an adaptation of existing scoring-rule-based loss functions that accommodates the use of an approximate Bayesian computation sequential Monte Carlo (ABC-SMC) framework. The fitted model achieves results comparable to those of two leading impact estimation tools in the prediction of total mortality when tested on a set of held-out past events. The proposed method provides four advantages over existing empirical approaches: modelling produces a gridded spatial map of the estimated impact, predictions benefit from the Bayesian quantification and interpretation of uncertainty, there is direct handling of multi-shock earthquake events, and the use of a joint model between impact types allows predictions to be updated as impact observations become available.} 


\maketitle

\section{Introduction}
In the first hours and days following a disaster event, such as an earthquake, a number of critical decisions are made in the coordination of disaster response efforts. These include decisions regarding the locations of emergency infrastructure \citep{nappi2019}, the amount of funding provided by major humanitarian donors \citep{OCHA2017}, and the mobilisation of emergency aid \citep{apte2010}. To best inform these decisions, governments and non-governmental organisations require early and accurate estimates of the crisis extent. A number of tools have been developed to provide such estimates, typically combining real-time information on the shaking intensity with existing data describing the exposed population and infrastructure. Uncertainty is present at numerous stages throughout this modelling process, not just via the population exposure, shaking intensity, and vulnerability data, but also through measurement error and missingness in past impact data used to train the model.  This motivates the development of robust statistical approaches that are best able to accommodate and quantify this uncertainty.

Many existing impact estimation tools adopt an analytical or engineering-based approach, in which comprehensive datasets describing the exposed infrastructure are combined with pre-specified relationships that model their response. These ``fragility curves'' are a central feature of most analytical earthquake impact models, describing the probability of a building meeting or exceeding a damage state at a given shaking intensity, and varying based on factors such as building material and the number of storeys \citep[e.g.,][]{rosti2021, khaloo2016}. The use of fragility curves is central to the Hazard US (HAZUS) model, which uses a collection of curves for 36 different building types and various other physical infrastructures, ranging from water storage tanks to airport terminals \citep{kircher2006}. 

Other leading tools that employ a more analytical approach include the Global Earthquake Model (GEM), the European Seismic Risk Model (ESRM20), and CLIMADA \citep{crowley2021, yepes2016,aznar2019}. The use of fragility curves, and the analytical approach more broadly, encounter two main limitations. First, the approach requires a complete taxonomy of building stock in the modelled region, which is unavailable or outdated in many locations. Second, when attempting to estimate the humanitarian impact, there is limited data available for modelling the relationship between building collapse and subsequent mortality or displacement. For example, ESRM20 fixes the probability of entrapment given collapse and subsequent deaths using a study with a sample size of 50 collapsed buildings, and does not include uncertainty in these probabilities \citep{crowley2021,reinoso2018}.

Other leading impact estimation tools employ a more empirical method, in which observations from past disasters are used to estimate statistical relationships that can be used for future prediction. In the empirical model presented as part of the Prompt Assessment of Global Earthquakes for Response (PAGER) system, a log-normal cumulative density function (CDF) is fitted between shaking intensity and fatality rate in each country \citep{jaiswal2010}. This method struggles, of course, when applied to countries for whichthere is little training data. For example, the fitted model for Argentina extrapolates poorly to high intensity events, suggesting that the mortality probability remains below 0.001 until the intensity is larger than 12.5 (note that the MMI scale describes an intensity of 12 as causing total damage). The Global Disaster Alert and Coordination System (GDACS) overcomes this limitation by specifying a single relationship between intensity and exposure, which is then rescaled according to country-specific vulnerability data \citep{gdacs}. This part of the GDACS model is hard to evaluate, and we cannot find publicly available information describing which parameters in GDACS are estimated statistically, and by what method. A different empirical model is developed by \citet{kam2024} for predicting human displacement caused by tropical cyclones. This provides spatial estimates that consider uncertainty in the impact function relating wind speed with the proportion of people displaced. \citet{kam2024} adopt an optimisation approach to fit a separate impact function to each past event, and then use the ensemble of resulting functions to represent uncertainty. 
Such an approach does not account for the uncertainty in each fitted function, which may be large when fitting a full curve based on a single total impact. In the extreme case, if there is only one past event on which to train the model, it creates the illusion that there is no uncertainty. 

In this work we employ the empirical rather than the analytical approach. This has the benefit of sharing information across the various impact types and affected regions, enabling us to update predictions according to the inferred conditional distributions as new observed data becomes available. For example, what is the updated prediction for population displacement after we have observed that the total mortality is larger than 3000? A second benefit of an empirical approach is that it facilitates uncertainty quantification, in our case via a general Bayesian framework. Bayesian inference has been applied across various aspects of disaster impact modelling, such as developing functions for mortality rate due to building collapse \citep{noh2017}, improving remote damage assessments from satellite images \citep{booth2011}, and updating PAGER estimates using incoming impact data \citep{noh2020}, but has not yet been used to construct a full impact estimation tool. Under the Bayesian interpretation of statistical estimates, we are able to answer questions within a framework that is more intuitive and well-suited for decision-making, as it gives direct access to estimates of probabilities of outcomes. For example,  the probability that an event's mortality exceeds a certain threshold. We are also able to quantify parameter uncertainty, and to propagate this uncertainty to predictions. 

To address the challenge of data scarcity often encountered by empirical models, we train a single global model and aim to capture differences in local vulnerability via relevant covariates, such as gross sub-national income per capita and the frequency of earthquakes in the region. In this way, predictions in countries with few historical events can still leverage information from countries with extensive past impact data. This approach also allows modelling to be performed at a higher resolution, in our case dividing the exposure into grid cells of 2.5 $\times$ 2.5 arc-minutes, resulting in disaggregated spatial impact estimates. 

Fitting such a model in a statistically rigorous way presents a number of challenges. Impact data is rarely available at the resolution at which modelling is performed, so calculation of the likelihood requires integrating over all possible realisations at the grid cell level, which is computationally infeasible. Also, standard likelihood-free approaches, such as approximate Bayesian computation, would typically require summarising over model samples from different events, which does not seem sensible given events can have vastly different intensities, exposures and vulnerabilities. To address these challenges, we use a scoring-rule based approach to generalized Bayesian likelihood-free inference, as proposed in \citet{pacchiardi2024}. Our work adapts this idea via the use of a sequential Monte Carlo (SMC) approximate Bayesian computation (ABC) algorithm, and a threshold-based adjustment to the scoring rule loss function. We find the ABC-SMC algorithm yields similar results to an ABC Markov chain Monte Carlo (MCMC) algorithm, but provides some benefits such as a larger scope for parallelisation. 

We name the developed impact estimation tool the Oxford Disaster Damage Real-Time Information Network (ODDRIN). In this work, we validate the algorithm using simulated earthquake events and impact data, and also present the results on a set of past earthquake events. The model is trained on aggregated impact data collected from various databases and reports, and we additionally explore the use of point building damage data and some of the associated weaknesses. We evaluate performance of the fitted model over a set of held-out testing events, and compare the predictions of total mortality to two leading impact estimation tools, GDACS and PAGER. While performance is comparable, the methodology employed presents a number of additional benefits over existing approaches. First, ODDRIN provides a gridded spatial map over the predicted impact, which has previously not been available for empirical models. Second, the full model is trained within a Bayesian framework and predictions benefit from the associated uncertainty quantification. Third, to the authors' knowledge, ODDRIN is the first earthquake impact estimation tool with a model that handles foreshocks and aftershocks. Fourth, a joint model is employed over various impact types, producing conditional distributions that permit prediction updates as incoming information becomes available. 

\section{Model and Methods}\label{sec:ModelMethods}
\subsection{Model}\label{subsec1}

\begin{table}
\centering
\begin{tabular}{ l l }
 \hline
 \textbf{Symbol} & \textbf{Definition} \\
 $I$ & 
    The shaking intensity using the modified Mercalli intensity\\
 $I_0$ & 
    The threshold intensity below which we assume damage is negligible \\
 $i$ & 
    Subscript to distinguish between hazard instances in the same event \\
 $j$ & 
    Subscript to distinguish between exposed grid cells \\
 $r$ & 
    Used to denote a region, such that $ j\in r$ means grid cell $j$ falls in region $r$ \\
 $q$ & 
    Subscript to distinguish between income quantiles \\
 $q_{(a,b)}$ & 
    As above, but specificially referencing the quantile containing the population from income \\
    & percentiles $a$ to $b$ \\
 $C_{(a,b)}$ & 
    The total share of the pretax national income earned by $q_{(a,b)}$ \\
 $\beta_1, \beta_2, \beta_3, \beta_4, \beta_5, \beta_6, \beta_7, \beta_8 $ & 
    Coefficients for the covariates \verb|Vs30|, \verb|PopDens|,  \verb|SHDI|,  \verb|GNIc|,  \verb|EQFreq|,  \verb|FirstHaz|,  \verb|Night| and \\
    & (\verb|FirstHaz| $\times$ \verb|Night|) respectively \\
 $D$ & 
    A latent variable that reflects an increasing function of damage, if the exposure was held constant \\
 $\mu_\text{Mort}, \mu_\text{Disp}, \mu_\text{BuildDam}$ & 
    Means of the normal cumulative density functions (CDF) linking the latent damage $D$ with the \\ 
    & probabilities  of mortality, displacement, and building damage respectively \\
$\kappa_\text{Mort}, \kappa_\text{Disp}, \kappa_\text{BuildDam}$ &
    Standard deviations of the normal CDFs linking the latent damage $D$ with the probabilities of\\ 
    & mortality, displacement, and building damage respectively \\
$\epsilon$ & 
    Local (grid cell level) stochastic component. \\
$\xi$ & 
    Event-wide stochastic component. \\
$\sigma_{\text{Mort}}, \sigma_{\text{Disp}}, \sigma_{\text{BuildDam}}$ & 
    Standard deviations of the marginal distributions for each impact type in $\xi$ \\
$\rho$ & 
    Correlation between impact types specified via the covariance of $\xi$ \\ 
$\tau$ & 
    The multiplicative factor between the covariance of $\xi$ and $\epsilon$ \\ 
$\sigma_{\text{Local}_{\text{Mort}}}$ & 
    The standard deviation for the marginal distribution for mortality in  $\epsilon$ \\ 
$\theta \in \Theta$ & 
    The set of model parameters \\
$N$ & 
    The number of events\\
$y$ & 
    Observed data\\ 
$x$ & 
    Data sampled from the model \\ 
$\eta(\cdot)$ & 
    Summary statistics over the data \\ 
$\lambda(\cdot)$ & 
    Distance function comparing observed data with model samples \\ 
$M$ & 
    The number of repeated model samples generated for each calculation of the distance function $\lambda()$ \\ 
$\omega$ & 
    The learning rate in a generalised Bayesian posterior \\ 
$\delta$ & 
    The upper threshold on the distance accepted under ABC \\ 
$\alpha$ & 
    Hyperparameter controlling the targeted reduction in effective sample size between steps in an \\
    & ABC-SMC algorithm \\ 
 \hline
\end{tabular}
 \caption{Notation used in describing the model and method}
 \label{table:notation_table}
\end{table}

Disaster impact models typically consist of three main components: the hazard, the vulnerability, and the exposure (e.g. exposed population). Following the practice of the disaster risk modelling community, we use the term hazard to describe the earthquake shock, and a disaster event consisting of multiple shocks will therefore consist of multiple hazards. The term hazard is therefore used even when describing shocks that have already occurred, differing from other fields in which hazard may be more commonly associated with potential or future harm. We work with the modified Mercalli intensity (MMI), which is available for earthquake events globally via `ShakeMaps' produced by the United States Geological Survey (USGS) \citep{wald2005}. Modelling across a spatial grid of the exposed region, we denote the MMI of hazard $i$ in spatial grid cell $j$ using $I_{i,j}$. To improve readability we only distinguish in notation between hazards within an event and not the events themselves, however, this dependence is implied throughout. A list of the notation used throughout this section can be found in Table \ref{table:notation_table}. 

For the vulnerability, we disaggregate the population into income quantiles, and for quantile $q$, apply a linear regression on various covariates to express the vulnerability, 
\begin{align}
    \text{Vulnerability}_{i,j,q} &= \beta_{1} \log(\texttt{Vs30}_{j}) + \beta_{2} \log(\texttt{PopDens}_{j}+0.1) + \beta_{3}\texttt{SHDI}_{j} + \beta_{4}\texttt{GNIc}_{j, q} + \nonumber \\
&\quad \beta_{5} \log(\texttt{EQFreq}_{j}+0.001) + \beta_{6}\texttt{FirstHaz}_{i} + \beta_{7}\texttt{Night}_{i} + \beta_{8} (\texttt{FirstHaz}_{i} \times \texttt{Night}_{i}), \label{eqn:vulnerability}
\end{align}
where each covariate (or transformed covariate) has been standardised to have mean zero and standard deviation one. We include two variables related to the hazard, $\verb|FirstHaz| \in \{0,1\}$ indicating whether the hazard is preceded by foreshocks, and $\verb|Night| \in \{0,1\}$ indicating whether the hazard occurs between 10pm and 6am local time. These terms are included as earthquakes at night are often modelled to have a larger mortality \citep[e.g.,][]{crowley2021}, and we expect this effect may be weakened if the exposed population is evacuated from buildings due to foreshocks. We therefore also include an interaction term between the two variables.  We include the relative earthquake frequency (\verb|EQFreq|) to account for increased preparedness in earthquake-prone regions, the subnational human development index (\verb|SHDI|) and gross sub-national income per capita (\verb|GNIc|) due to their relationship with physical, economic and social infrastructure, the population density (\verb|PopDens|) to account for how effects may vary between rural and urban areas, and the time-averaged shear-wave velocity to 30m depth (Vs30) as a proxy for soil stiffness. The latter variable is already included in the intensity estimation performed by USGS ShakeMaps, however, it may be also related to the impact via other mechanisms (e.g. liquefaction), and is included as a covariate in similar work \citep[e.g.,][]{loos2022}. There are a number of other variables that could be included, such as the average building age or predominant building material, however, it is challenging to find open-source and standardised global databases for this information.

The division of the population into income quantiles is to better reflect inequality that may exist beyond the subnational \verb|GNIc|. If $q_{(a,b)}$ refers to the quantile containing the population from percentile $a$ to $b$, and $C_{(a,b)}$ denotes this quantile's share of the pretax national income, then the \verb|GNIc| in each quantile is given by
\begin{equation}
\text{GNIc}_{j, q_{(a,b)}} = \text{GNIc}_{j} \times C_{(a, b)}  \times \frac{100}{b-a} \;.
\end{equation}
Given that the \verb|GNIc| is available sub-nationally, but the income quantile (income inequality) data only nationally, combining the extremes of each will inflate the true inequality. As a crude workaround, we disregard the income shares $C_{(0, 10)}$ and $C_{(90, 100)}$ and work with the remaining eight deciles. We therefore instead calculate 
\begin{equation}
\text{GNIc}_{j, q_{(a,b)}} = \text{GNIc}_{j} \times C_{(0.8a+10, 0.8b+10)}  \times \frac{100}{(0.8b+10)-(0.8a+10)} \;, \label{eq:GNIbreakdown}
\end{equation}
which compacts the income shares used and avoids the outer values. For example, the population in the quantile from percentiles 0 to 12.5 is given the income share of the population from percentiles 10 to 20, scaled appropriately for the sizes of the quantiles. 

For each impact type, $\text{Impact} \in \{\text{Mort}, \text{Disp}, \text{BuildDam}\}$, we combine the hazard and vulnerability components to obtain a latent variable $D$,
\begin{equation}
    D_{\text{Impact},i,j,q} = (I_{i,j} + \text{Vulnerability}_{i,j,q} + \epsilon_{\text{Impact},i,j} +  \xi_{\text{Impact}}) \times \mathds{1} [I_{i,j} \geq I_0] \label{eqn:d_latent}
\end{equation}
 For computational purposes, we introduce an intensity threshold $I_0$ below which we assume the likelihood of damage to be negligible, allowing modelling to be contained to regions where $I > I_0$. In practice, we set $I_0=4.3$, as we found during data collection that very little damage occurs where the intensity is less than 4.3. Note that when calculating $D_{\text{BuildDam}}$, we do not disaggregate by income quantile and calculate the vulnerability using the median \verb|GNIc| value. Equation \ref{eqn:d_latent} includes a local stochastic component $\epsilon_{\text{Impact}, i,j}$ and an event-wide stochastic component $\xi_{\text{Impact}}$. The former varies between each grid cell, hazard instance (e.g. principal and aftershocks) and impact type, while the latter varies between impact types and events but remains constant over all grid cells and hazard instances in an event. These terms aim to represent not just inherent randomness, but also data error and uncertainty not captured by the model, such as the ShakeMap underestimating the intensity at a specific grid cell, or the buildings exposed to a particular event being older on average. The marginal distribution of each component in the event-wide error term is drawn from a normal distribution with mean $0$ and variance $\sigma^2_{\text{Impact}}$, with a correlation $\rho$ introduced between impact types in the joint distribution. It is likely the case that the correlation is not identical between all impact types, however, exploratory data analysis suggests that it is sufficiently similar to avoid the burden of two additional parameters. We therefore have
\begin{equation}
\begin{bmatrix}
    \xi_{\text{Mort}} \\
    \xi_{\text{Disp}} \\
    \xi_{\text{BuildDam}}
\end{bmatrix} \sim N\left(\begin{bmatrix} 0 \\ 0 \\ 0
\end{bmatrix}, \begin{bmatrix}
    \sigma^2_{\text{Mort}} & \rho \sigma_{\text{Mort}} \sigma_{\text{Disp}} & \rho \sigma_{\text{Mort}} \sigma_{\text{BuildDam}} \\
    \rho \sigma_{\text{Mort}} \sigma_{\text{Disp}} & \sigma^2_{\text{Disp}} & \rho \sigma_{\text{Disp}} \sigma_{\text{BuildDam}} \\
    \rho \sigma_{\text{Mort}} \sigma_{\text{BuildDam}} & \rho \sigma_{\text{Disp}} \sigma_{\text{BuildDam}} & \sigma^2_{\text{BuildDam}}
\end{bmatrix} \right). 
\end{equation}
For the local error term, we also use a multivariate normal distribution with zero mean vector. Practically, we have insufficient observation data at the grid cell level to perform meaningful inference about the full covariance structure, but we expect the correlation and relative variance of each impact type to be similar to the event-wide error. We therefore use the same covariance matrix but multiply by scale parameter $\tau$. However, rather than modelling $\tau$, which is strongly correlated to the terms comprising the event-wide error covariance matrix, we instead model the resulting standard deviation for mortality in the local error, $\sigma_{\text{Local}_\text{Mort}} = \sqrt{\tau}\sigma_{\text{Mort}}$. 

The term $D_{\text{Impact}, i, j, q}$ has no direct physical interpretation but, it is an increasing function of the damage for each fixed exposure. We therefore transform it to a probability between $0$ and $1$, allowing it to be related to the exposure, with $p^{\text{Mort}}$ and $p^{\text{Disp}}$ representing the probability of a person dying and becoming displaced respectively, and $p^{\text{BuildDam}}$ representing the probability of a building being damaged or destroyed. 
Using $\Phi(\cdot | \mu, \sigma)$ to denote the normal CDF with mean $\mu$ and standard deviation $\sigma$,
\begin{align}
p^{\text{Mort}}_{i,j,q} &= \Phi(D_{\text{Mort},i,j,q} \mid \mu_{\text{Mort}}, \kappa_{\text{Mort}}), \label{eqn:mortimpact} \\
p^{\text{Disp}}_{i,j,q} &= \max \{ \Phi(D_{\text{Disp},i,j,q} \mid \mu_{\text{Disp}}, \kappa_{\text{Disp}}) -p^{\text{Mort}}_{i,j,q}, 0 \}, \label{eqn:dispimpact}\\
p^{\text{BuildDam}}_{i,j} &= \Phi(D_{\text{BuildDam},i,j} \mid \mu_{\text{BuildDam}}, \kappa_{\text{BuildDam}}). \label{eqn:bdimpact}
\end{align}
 The construction of $p^{\text{Disp}}$ ensures that the probabilities of displacement and mortality do not sum to larger than one (a person cannot be both displaced and deceased). The choice to use a normal CDF to produce impact probabilities disagrees with some the existing literature, which typically employs the log-normal CDF. However, models using the log-normal distribution have previously exhibited underfitting at high intensities in order to capture the gradual increase in impact probability over lower intensities \citep[e.g.,][]{jaiswal2010}, an issue which is exacerbated by the right skew of the log-normal distribution. The use of the log-normal CDF is better established for building-fragility curves, however, these are usually as a function of peak ground acceleration rather than intensity, which are in turn roughly exponentially related \citep{wald1999}. The normal CDF is also closer to a log-linear model, which was shown to perform strongly in \citet{firuzi2020} and \citet{li2021} but is not constrained to be less than one. 
 There are other decisions made throughout the construction of the model, such as the adoption of an additive form in Equations \ref{eqn:vulnerability} and \ref{eqn:d_latent}, and the use of normally distributed errors. Given the challenge of performing exploratory data analysis when working with a latent model layer, these choices typically reflect default statistical practice and aim to produce the most interpretable model. Experimentation with replacing $I_{i,j}$ with $\exp(\gamma I_{i,j})$, where $\gamma$ is a parameter to be estimated, was not found to improve performance, but led to identifiability challenges that slowed model fitting. 

Finally, we construct a model for the observed impact by combining the impact probabilities with the exposure. The displaced population $\text{Disp}_{i,j,q}$, the deceased population $\text{Mort}_{i,j,q}$, and the remaining population neither displaced nor deceased $\text{Rem}_{i,j,q}$ follow the model
\begin{equation} \label{eq:PopMultinom}
(\text{Disp}_{i,j,q}, \text{Mort}_{i,j,q}, \text{Rem}_{ij,q}) \sim \Multinomial \left(\text{Pop}_{0,j,q}, (p_{i,j,q}^{\text{Disp}},p_{i,j,q}^{\text{Mort}},1-p_{i,j,q}^{\text{Disp}}-p_{i,j,q}^{\text{Mort}})\right),
\end{equation}
where $\text{Pop}_{0,j,q}$ is the initial population. Similarly, the number of buildings damaged $\text{BuildDam}_{i,j}$ is modelled using a binomial distribution,
\begin{equation} \label{eq:BuildDam}
\text{BuildDam}_{i,j} \sim \Binom \left(\text{Build}_{0,j} \;, \; p_{i,j}^{\text{BuildDam}} \right) \\
\end{equation}
where $\text{Build}_{0,j}$ is the original number of buildings. In the case of an earthquake with multiple shocks, the model is applied to each hazard one-by-one with impact probabilities calculated using each hazard's intensity. For all hazards except the first, we replace $\text{Pop}_{0,j,q}$ in Equation \ref{eq:PopMultinom} with the remaining population unaffected after the previous hazard, and $\text{Build}_{0,j}$ with the number of undamaged buildings remaining.

We do not model the transition from displaced to deceased (for example, due to illness in an emergency shelter) and instead seek mortality figures that reflect casualties caused directly by the hazards. Following all hazards, the total impact in region $r$ is calculated by
\begin{align}
\text{Mort}_r &= \sum_{j \in r} \sum_{i} \sum_{q} \text{Mort}_{i, j, q} \;, \\
\text{Disp}_r &= \sum_{j \in r} \sum_{i} \sum_{q} \text{Disp}_{i, j, q} \;, \\
\text{BuildDam}_r &= \sum_{j \in r} \sum_{i} \text{BuildDam}_{i, j} \;,
\end{align}
or a gridded spatial impact prediction obtained by simply summing over hazards and quantiles. 

\subsection{Method} \label{sec:Method}
\subsubsection*{Likelihood-Free Bayesian Inference}

Denoting observed data by $y_{1:N} \in \mathcal{Y}^N$ and parameters by $\theta \in \Theta$, explicit calculation of the likelihood $p(y_{1:N}|\theta)$ is not practical for this model. Since much of the observed impact data is aggregated, obtaining the likelihood would require summing over the likelihood of all possible grid cell realisations that total to the aggregated values, which is computationally infeasible. We instead use a likelihood-free approach, in which a given set of parameters is evaluated by the similarity between the observed data and a sample from the model under this parameterisation. For simulated data $x_{1:N}$, a basic likelihood-free posterior is given by:
\begin{equation}
p_{\mathrm{LF}}(\theta, x_{1:N}  \mid y_{1:N}) \propto \mathds{1} \left[\lambda \left( \eta(y_{1:N}), \eta(x_{1:N})\right) \leq \delta \right] p(x_{1:N} \mid \theta) p(\theta), \label{eq:abc_post}
\end{equation}
where $\lambda(\cdot)$ is a distance function comparing summary statistics $\eta(\cdot)$ of the simulated data and observed data, and $\delta$ provides an upper threshold on this distance. We use $n=1, \dots, N$ to index the event, therefore each $y_n$ and $x_n$ may be multivariate and comprise multiple observations across different locations and impact types. The purpose of the summary statistics is to reduce the dimensionality of the data; if the number of data points is too large then it is very unlikely that a sample will closely match the observations across every data point, even if they are drawn from the same distribution. However, unless sufficient statistics are available, summary statistics may not capture all features of the data, therefore introducing a trade-off between information loss and dimensionality. While there have been numerous methods developed for automatically selecting summary statistics that aim to mitigate information loss \citep[e.g.,][]{prangle2014, jiang2017}, these approaches typically do not accommodate covariates, and instead require identically distributed observations. In our application, summarising over observations without consideration of the vastly different hazards, vulnerabilities, and exposures would result in significant information loss. 
We therefore turn instead to a scoring-rule approach similar to that of \citet{pacchiardi2024}. Here, we compare each observation $y_n$ to $M$ samples $x_{n, 1:M}$ by a scoring rule SR, and require that the average score fall below threshold $\delta$,
\begin{equation}
p_{\mathrm{LF}}(\theta, x  \mid y) \propto \mathds{1} \left[ \left(\frac{1}{N} \sum_{n=1}^N \mathrm{SR}(y_n, x_{n,1:M}) \right) \leq \delta \right] p(x \mid \theta) p(\theta), \label{eq:scoringrulepost}
\end{equation}
where $y=y_{1:N}$ and $x$ denotes the collection of all $M$ samples across the $N$ events, $x_{1:N, 1:M}$. As M approaches infinity, $\mathrm{SR}(y_n, x_{n, 1:M})$ becomes fixed given $\theta$ and the posterior for $\theta$ is proportional to the prior where $\frac{1}{N} \sum_{n=1}^N \mathrm{SR}(y_n, x_{n, 1:M}) < \delta$. In Equation \ref{eq:abc_post}, if the tolerance $\delta$ is set to zero and the summary statistics are sufficient, then we return the true Bayesian posterior. This is not the case in Equation \ref{eq:scoringrulepost}, which needs to be interpreted in a more general framework. Augmenting the generalised Bayesian posterior defined in \citet{bissiri2016} to include sampled data gives
\begin{equation}
p(\theta, x  \mid y) \propto \exp(- \omega \ell(\theta, y, x)) p(x|\theta) p(\theta). \label{eq:generalisedposterior}
\end{equation}
where $\omega$ is a parameter that controls the learning rate and $\ell(\theta,y,x)$ is a loss function. 
In Equation \ref{eq:scoringrulepost}, the loss function is therefore defined by
\begin{equation}
\ell(\theta, y, x) =  \begin{cases}
0, \qquad \left(\frac{1}{N} \sum_{n=1}^N \mathrm{SR}(y_n, x_{n, 1:M})\right) \leq \delta   \\
\infty , \qquad \text{otherwise.} 
\end{cases} \label{eq:lossfunction}
\end{equation}
The generalised posterior in \citet{bissiri2016} calls for additive loss functions, which is not the case in Equation \ref{eq:lossfunction}. However, the requirement for additivity is relaxed in \citet{knoblauch2022}, which accepts an infinite sequence of loss functions $\{L_N\}_{N \in \mathbb{N}}$ specified on the sequence of spaces $\{\Theta \times \mathcal{Y}^N \}_{N \in \mathbb{N}}$. 

\citet{knoblauch2022} adopt an optimisation-centric view of Bayes' Rule, and justify this approach by arguing that several principles traditionally supposed to underlie Bayesian inference are rarely upheld in modern practice. Standard Bayesian inference assumes correct specification of the likelihood, the ability of practitioners to fully convey prior believes via probability distributions, and the availability of infinite computational resources. All three of these ingredients are problematic in our setting: model misspecification arises from both measurement error and the use of a simplified model to represent complex real-world events, the interaction of parameters via a latent variable makes it challenging to construct prior distributions, and we cannot simulate infinite model samples $x$. 
There are several examples of non-additive loss functions adopted within a Bayesian framework \citep[e.g.,][]{cherief2020, hooker2014, prangle2017}, typically to improve robustness to model misspecification. 

For the scoring rule, we use the energy score, calculated by
\begin{equation}
\text{Energy Score}(x_{n,1:M}, y_n) = \frac{1}{M} \sum_{j=1}^{M}||x_{n,j} - y_n ||-\frac{1}{2M^2}\sum_{i=1}^{M}\sum_{j=1}^{M} || x_{n,i} - x_{n,j}||.
      \end{equation}
The primary motivation for use of the energy score is that it accommodates multivariate data, so models can be rewarded for having appropriately calibrated joint distributions among different impact types and spatial regions for the same event. We choose the energy score over other multivariate kernel scoring rules for convenience of interpretation; it reduces to the continuous ranked probability score in the univariate case, which in turn reduces to the absolute deviation when only a single sample is obtained. One associated challenge with the energy score is that both its expectation and variance change with the number of samples $M$. In particular, our experimentation shows that small $M$ favours underdispersed samples, as detailed in Appendix \ref{appendix:es_bias}. In practice, we therefore use $M \geq 60$, which also reduces the variance of the calculated energy scores and produces more stable results. A second property of the energy score is that it is sensitive to the coordinate system used. In our application, it seems sensible to accommodate larger errors as the scale of the impact increases; for example, we would like to penalise a difference between 0 and 50 more than between 5000 and 5050. We therefore work on a $\log(y+10)$ scale. The addition of a constant before applying the log function allows for zero values; the exact constant is somewhat arbitrary, but $10$ is large enough to avoid overemphasising small errors close to 0. 
With this scale, the observation-sample pairs $(0,5)$, $(100, 155)$, and $(10\;000, 15\;005)$, for example, all result in equal contributions to the distance function. In Appendix \ref{appendix:mse_loss} we compare the model fit using the energy score to that using the mean Euclidean distance, which underestimates the error terms and produces poorly calibrated posterior predictive distributions.

Given the number of parameters, a straightforward likelihood-free rejection scheme will be extremely inefficient or require a large tolerance for the distance between the observed and simulated data. We therefore employ a Sequential Monte Carlo (SMC) algorithm, wherein a sequence of sampling densities is generated that allows gradual progression of samples, referred to as particles, from the prior to the target, with the tolerance $\delta$ decreasing between each step. \citep{sisson2007sequential}. We use the approximate Bayesian computation (ABC) SMC algorithm described in  \citet{delmoral2012}, as the standard ABC-SMC algorithm can lead to particle degeneracy in high-dimensional spaces \citep{bengtsson2008}. SMC presents a number of benefits over alternatives, such as MCMC, making it popular in likelihood-free settings. A major benefit is that ABC-SMC is inherently parallel; particles can be divided across resources at each step and samples from the likelihood obtained independently. ABC-SMC also provides better handling of multi-modal posterior distributions, and allows adaptive selection of the tolerance $\delta$ (while loss functions such as that in Equation \ref{eq:generalisedposterior} do not include $\delta$, they instead require specification of learning rate $\omega$). While these benefits make it worth expanding the SR-posterior literature to include ABC-SMC, it is worth noting that there are also some drawbacks, such as difficulties in selecting the `stopping point' and a breakdown of the approximation when the difference between successive densities is too large. 

Within the ABC-SMC algorithm proposed in \citet{delmoral2012}, particles are perturbed at each step using an MCMC kernel with a Metropolis--Hastings acceptance ratio, and are resampled from the current population when the effective sample size falls below a certain threshold. To avoid discarding too many particles between steps, the decrease in tolerance is determined adaptively according to a hyperparameter $\alpha$, where $1-\alpha$ is the targeted decrease in effective sample size between successive steps. This does not fully safeguard against particle degeneracy: if the acceptance rate of the MCMC kernel is too low, then particles may still be discarded and resampled faster than they can be perturbed, which eventually leads to many duplicates of the same particle. To address this, we adaptively select $1 - \alpha$ equal to the acceptance rate of the perturbations at the previous step. Therefore, the number of duplicates made at the resampling step should be roughly similar to the number of particles perturbed by the next resampling step. This heuristic ignores that some particles may be perturbed multiple times and others not at all; however, we have found that this approach works better in practice than holding to a fixed $\alpha$ when the acceptance rate is low, particularly in the later stages of the algorithm. For the perturbation kernel, we calculate the optimal multivariate normal perturbation kernel covariance described in \citet{filippi2013} and divide by 5. Initial runs of the algorithm showed that the original covariance led to a very low acceptance rate (no accepted proposals over the first 5 steps) while dividing by larger values leads to particles not being able to effectively explore the parameter space, resulting in clusters of particles. In Sections \ref{sec:Testing} and \ref{sec:Results}, we produce results using $1000$ ABC-SMC particles and resample when the effective sample size falls below $500$. We calculate the energy score using $M=100$ samples but have no additional pseudo-marginal layer; that is, we calculate the energy score only once per particle. 

\subsubsection*{Measurement Error} \label{sec:measurement_error}
To account for measurement error, we also adjust the relative contribution to the energy score of the different impact types. The tolerance $\delta$ represents the discrepancy permitted, on average, between a distribution of samples and the corresponding observation, so it is sensible to permit larger discrepancies where the observation itself is subject to a high degree of uncertainty. While uncertainty information about specific observations is generally unavailable, we do have intuition about the relative uncertainty of different impact types, for example, displacement data is typically very unreliable whereas observations of deaths are more accurate due to regulation of death certification. We note, however, that death certification accuracy and completeness varies from country-to-country. To estimate the relative uncertainties of the different impact types, we take the events where different data sources provide observations for the same impact and region, and calculate the mean absolute deviation of each after applying the $\log(y+10)$ transformation, then average across impact types. The mean energy score represents a multivariate and probabilistic extension of the mean absolute deviation, so this compares observations in a similar way to the distance function employed. Taking the inverse of the averaged mean absolute deviations and normalising so that the displacement weight is $1$ yields weightings of $7$ and $0.6$ for mortality and building damage respectively. These are based on 74 data points for mortality where at least two different sources provide observations, 23 data points for displacement, and 35 for building damage. The weights are implemented in the energy score by multiplying $\log(y+10)$ by the weight before applying the energy score, thus weighting their relative contributions. Under these choices, a simulated mortality of 95, a simulated displacement of 75, and a simulated building damage of 55 are all roughly `equidistant' from respective observations of 100. This approach aims to penalise based on difference from the possible underlying truth rather than just the observed value. 

\subsubsection*{Prior Elicitation}
We construct prior distributions over the parameters by introducing prior beliefs in two stages. First, we select a prior distribution for each parameter; these priors are largely non-informative although most bound the support of the parameters based on their physical interpretation. Second, to reflect prior beliefs about combinations of parameters, we further constrain the support by assigning a prior weight of zero to parameter transformations that produce unreasonable impact probabilities for given intensities.

Within the first stage, rather than eliciting informative priors through expert knowledge, we instead opt for relatively non-informative priors that are constrained only to eliminate unplausible physical relationships. Under this framework, a uniform prior with bounds outlined in Table \ref{table:unifp_bounds} is applied to all parameters except the vulnerability covariate coefficients. To explain the bounds chosen, we begin with the parameters $\mu_\text{Mort}$ and $\kappa_\text{Mort}$. Here, consider an exposed grid cell with standardised vulnerability terms and stochastic terms at their means of zero. We therefore have $D_{\text{Mort},i,j,q}=I_{i,j}$ and
\begin{equation} \label{eq:norm_priors}
p^{\text{Mort}}_{i,j,q} = \Phi(I_{i,j}|\mu_\text{Mort}, \kappa_\text{Mort}).
\end{equation}
Therefore, the parameter $\mu_\text{Mort}$ represents the MMI that results in a fatality probability of $50\%$.
Given the physical ground conditions associated with different shaking intensities, we constrain this to be between 9 and 13.5. For the same grid cell, using the quantiles of the normal distribution, a fatality probability of $0.135\%$ will occur at an intensity of $\mu_\text{Mort} - 3\kappa_\text{Mort}$. We therefore bound $\kappa_\text{Mort}$ to reflect the change in intensity over which this reduction in probability is plausible. At the upper bound of $3$ of the prior for $\kappa_{\text{Mort}}$, moving down from an impact probability of 50\% to 0.135\% would require a decrease in 9 of the intensity, while at the lower bound, 0.25, this would take an intensity decrease in 0.75. It is clear that the true value would lie somewhere between these two extremes. We similarly derive the priors for $\mu_\text{Disp}$, $\kappa_\text{Disp}$, $\mu_\text{BuildDam}$, and $\kappa_\text{BuildDam}$.

To determine appropriate priors for the vulnerability and stochastic terms, we consider reasonable constraints for their effect relative to changes in the shaking intensity. Considering an exposed grid cell with all vulnerability terms at zero and $I>I_0$, we have
\begin{equation}
D_{\text{Mort},i,j,q} = I_{i,j} + \xi_{\text{Mort}} + \epsilon_{\text{Mort},i,j}.
\end{equation}
Therefore, varying the value of the local or event-wide stochastic terms has the same effect as increasing or decreasing the intensity by the same amount. For example, if $\epsilon_{\text{Mort}, i,j}$ equals zero in one grid cell and four in another, then this is equivalent to the latter grid cell being exposed to an intensity larger by four (with all else held constant). We set the upper bound of $\sigma_{\text{Local}_{\text{Mort}}}$ at $2$ as this has $2.5\%$ and $97.5\%$ percentiles of $-4$ and $4$, and set the upper bounds of $\sigma_\text{Mort},\sigma_\text{Disp}$, and $\sigma_\text{BuildDam}$ at 1.5 as this produces $2.5\%$ and $97.5\%$ percentiles of $-3$ and $3$. These upper bounds seem generous given the the range from a `light' shaking intensity (4 MMI) to an `extreme' shaking intensity (10 MMI).
For each of the vulnerability covariate coefficients $\beta$, we use a Laplace prior centered at 0 with scale parameter 0.2. The 2.5\% and 97.5\% quantiles for this distribution lie at around -0.6 and 0.6 which, when combined with extremes of the standardised vulnerability covariates (taking values around -3 or 3), becomes equivalent to moving the intensity down or up by around 1.8. The use of the Laplace distribution produces a more informative prior than if a uniform distribution was used, leading to more conservative parameter estimates that are less prone to overfitting.

\begin{table}
\centering
\begin{tabular}{ l r r }
 \hline
 Parameter & Lower Bound & Upper Bound \\
 $\mu_\text{Mort}$ & 9 & 13.5 \\
 $\kappa_\text{Mort}$ & 0.25 &  3 \\
 $\mu_\text{Disp}$ & 6.5 & 10.5 \\
 $\kappa_\text{Disp}$ & 0.25 & 3 \\
 $\mu_\text{BuildDam}$ & 6.5 & 10 \\
 $\kappa_\text{BuildDam}$ & 0.25 & 3 \\
 $\sigma_{\text{Mort}}$ & 0 & 1.5 \\ 
 $\sigma_{\text{Disp}}$ & 0 & 1.5 \\ 
 $\sigma_{\text{BuildDam}}$ & 0 & 1.5 \\ 
 $\sigma_{\text{Local}_\text{Mort}}$ & 0 & 2 \\ 
 $\rho$ & 0 & 1 \\ 
 \hline
\end{tabular}
 \caption{Lower and upper bounds of the prior distribution.}
 \label{table:unifp_bounds}
\end{table}

In the second stage of implementing the prior, we use the parameterised model to calculate the probability of mortality, displacement, and building damage at a low, medium and high hazard intensity, and test whether the impact probabilities fall within reasonable bounds (outlined in Table \ref{table:hlp_bounds}). We refer to this as a `higher-level prior', however, it is equivalent to constraining the support of a transformed version of the parameters. When performing the higher-level prior testing, we set the local and event-wide stochastic components to zero, as well as the standardised vulnerability covariates. Parameterisations that result in any probabilities outside the bounds are assigned a prior probability of zero. In cases where the higher-level prior is not satisfied, the proposed parameters can be immediately rejected without sampling from the likelihood, thus circumventing the most computationally demanding step in the algorithm. Note that for the results on the simulated data outlined in Section \ref{sec:Testing}, we also include the vulnerability terms in the higher-level prior, as detailed in Appendix \ref{appendix:HLPrior_SimData}. However, they were removed when fitting the real data, as it was found that the approach eliminated parameterisations with otherwise high posterior probability. 

\begin{table}
\centering
\begin{tabular}{ r @{\hskip 15pt} r @{\hskip 15pt} r @{\hskip 15pt} r @{\hskip 15pt} r }
 \hline
 \rule{0pt}{3ex} 
 Intensity & $p^{\text{Mort}}$ & $p^{\text{Disp}} + p^{\text{Mort}}$ 
 & $p^{\text{BuildDam}}$ & $p^{\text{Disp}}$  \\
 $4.6$ & (0,$10^{-6}$) & (0,0.01) & (0,0.05)  &  - \\ 
 $7$ & (0,0.01) & (0, 0.2) & ($10^{-6}$, 0.4)  & - \\ 
 $8$ & - & - & - & ($p^{\text{Mort}}$, 1)  \\ 
 $9.5$ & ($10^{-6}$,1) & (0.2,1) & (0.3,1) & -  \\
 \hline
\end{tabular}
 \caption{Bounds in the form (lower bound, upper bound) used in the higher level priors. Due to the relationship between the displacement and mortality probabilities, most checks are performed on their sum rather than the displacement probability itself, with an additional check that the probability of displacement is larger than the probability of mortality at an intensity of 8. }
 \label{table:hlp_bounds}
\end{table}

\section{Data} \label{sec:Data}

\begin{figure}[t]
\centering
\includegraphics[width=0.7\linewidth]{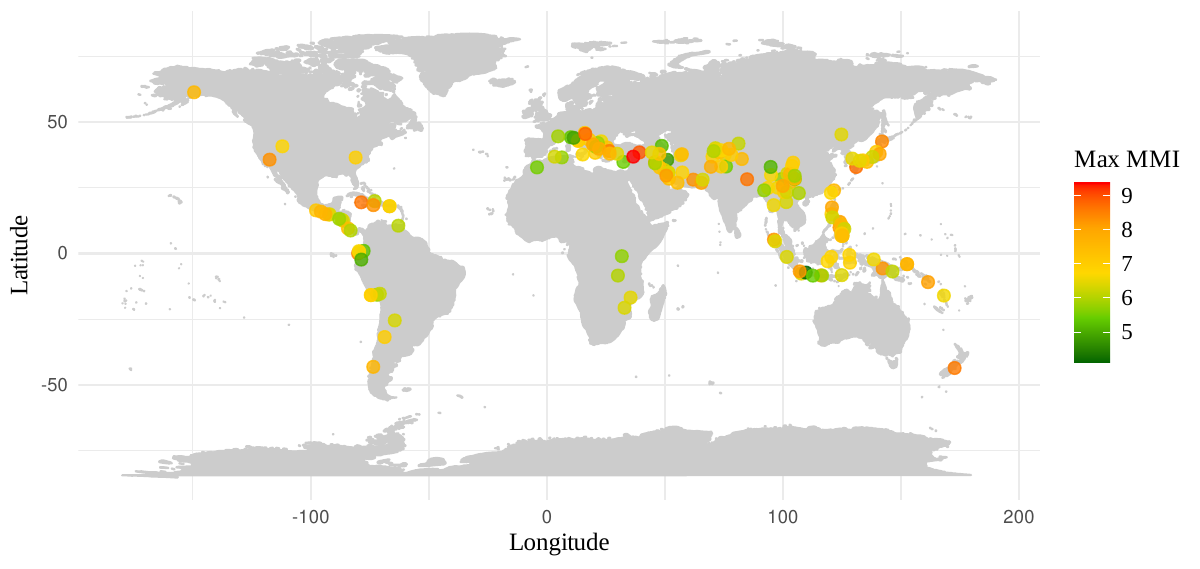}
\caption{The locations and maximum shaking intensities of the earthquake events used to train and test the model. }
\label{fig:event_map}
\end{figure}

The model is fit using a dataset of 167 real earthquake events spanning from 2011 to 2023, which we assembled using data from a diverse set of sources, as described in \citet{patten2024}. The events cover 76 countries and have maximum intensities ranging from 4.8 to 9.3. In Figure \ref{fig:event_map} we display the global distribution of the events together with their maximum intensities. The hazard data is obtained from USGS ShakeMaps, which interpolates over recorded ground motions to issue MMI estimates at a 30 by 30 arcsecond resolution. For most events, we include any foreshocks occurring within one week before the main shock and aftershocks within two weeks after. In cases where impact data is reported separately for two events, despite them occurring at the same location within this two week period, we separate them into different events. For computational purposes, we aggregate the ShakeMaps to a 2.5 by 2.5 arcminute resolution using bilinear interpolation. For the population data, we use the unconstrained yearly 1km resolution population data from WorldPop, which we aggregate to the resolution of the hazard data \citep{tatem2017}. Building count data is not yet available globally, but we use Bing Maps Global Building Footprints where available, which results in coverage of 889 observations across 50 of the events \citep{microsoft_building_footprints}. Note that although all observations represent aggregated building damage counts, an event can have multiple observations if the counts are aggregated at a subnational level. The Bing building footprints are produced using machine learning models trained on satellite imagery, and we generate the exposure data by counting the number of building footprints within each of the modelled grid cells. 

Vulnerability data is collected from a range of sources; the \verb|SHDI| and \verb|GNIc| data are obtained from Global Data Lab \citep{smits2019}, the \verb|Vs30| data is retrieved from USGS \citep{heath2020}, and the \verb|EQFreq| data is obtained from Global Earthquake Model and describes the peak ground acceleration with a 10\% probability of being exceeded in a 50 year return period \citep{johnson2023}. The income inequality data used to distribute the \verb|GNIc| between income quantiles is obtained from the World Inequality Database \citep{alvaredo2022}. The \verb|SHDI| and \verb|GNIc| data are available subnationally, with the level of aggregation varying between nations, the income inequality data is available nationally, and the \verb|Vs30| and \verb|EQFreq| data are available at a 30 arcsecond and 3 arcminute resolution respectively, with both aggregated to the resolution of the modelled grid cells using bilinear interpolation. 

The exposure, hazard, and vulnerability data are all subject to numerous sources of uncertainty. Population counts can exhibit daily, monthly and seasonal trends that are not captured via a single yearly figure; and the satellite imagery used to produce building count data is only known to be taken between 2014 and 2023, over which time the physical infrastructure may have varied significantly. The ShakeMaps are also subject to uncertainty, particularly in rural locations further from recording instrumentation. While USGS issues uncertainty values, it does not describe the error covariance structure between locations, which has a significant impact on prediction. The data for SHDI, GNIc and income inequality are not available at the resolution of the modelled grid cells, and the values used represent averages over larger areas in which there may be large disparities. These many sources of uncertainty motivate the inclusion of the local and event-wide error terms. 

\subsection{Aggregated Impact Data} \label{sec:AggImpactData}

Impact data is collected from a range of databases, articles and news reports, which we have compiled and made available\footnote{\url{https://figshare.com/projects/Hazard_vulnerability_exposure_and_impact_data_for_a_collection_of_past_earthquake_events/233999}}. National mortality data is predominantly collected from the Emergency Events Database (EM-DAT), displacement data from the Global Internal Displacement Database (GIDD) produced by the Global Internal Displacement Monitoring Centre (IDMC), and building damage and destruction data from the Significant Earthquake Database and the Earthquake Impact Database \citep{guha2009, IDMC, ncei2023, eqid2023}. In addition to national and total impact data, there is also subnational impact data available for many events. This is typically retrieved from news articles and reports by government agencies. However, if the subnational data is missing not at random (MNAR), that is, its selection is not random conditioned on the modelled covariates, then its inclusion has the potential to introduce bias. For example, if an event's subnational mortality is reported exclusively in regions where the mortality exceeds zero, then including only this data will bias the model. We address this challenge for the mortality data by checking that the subnational values available sum to the total mortality described by a different source, then inferring zeros only in remaining regions at the same administrative level as those with observations. Handling the displacement and building damage data is more challenging, as total impact values are often unreliable. We therefore address this event by event, keeping the subnational data when it appears to be missing at random, inferring zeroes in remaining regions when it appears that impact reporting is conditioned on non-zero values, and removing subnational data when it is unclear which approach to take.

Past impact data is subject to numerous sources of uncertainty. As mentioned previously, mortality data is typically the most accurate. However, there are some examples in which the data can be subject to uncertainty; figures provided by governments or NGOs may be untrustworthy, the mortality may be so large that the exact figure is unknown, or it may be unclear whether to attribute a fatality to the disaster event or other causes, such as subsequent illness or injury. Nonetheless, mortality figures provided by different sources usually agree. Measuring the displaced population is more challenging. While headcounts of those staying in emergency shelters are relatively straightforward to obtain, many of those displaced instead stay elsewhere, such as with family or friends, or remain outdoors, such as in tents or cars \citep{paul2023}. This is referred to as visible and invisible displacement. Although many methods seek to address this, such as household surveys or mobile phone data analysis, large uncertainties remain. For some events we have found the population displacement reported by different sources ranges from less than one hundred to several thousand. Building damage data is also subject to a large amount of measurement error. The definition of building damage may vary between events and, if assessed manually, is further subject to the judgement of those performing the evaluation. There are also examples of events in which only some exposed towns or areas are assessed for building damage, leading to underreporting of the true impact. As discussed in Section \ref{sec:Method}, we account for the relative uncertainties of each impact type by weighting their contribution to the energy score. 

\subsection{Point Impact Data} \label{sec:PointImpactData}
\begin{figure}[t]
\centering
\includegraphics[width=0.8\linewidth]{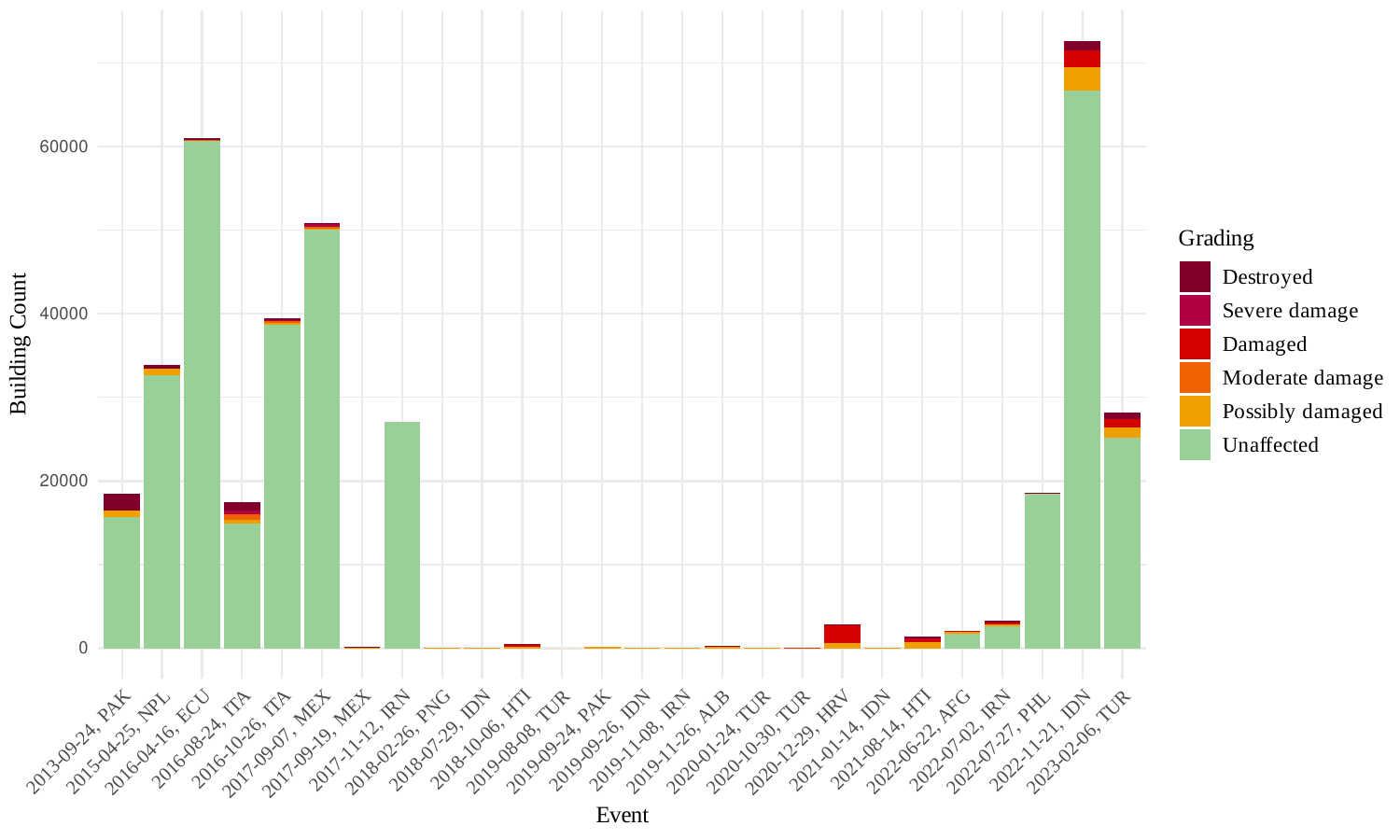}
\caption{The number of buildings classified under each of the damage gradings across the 26 events for which point damage classification data is available.}
\label{fig:PointSummary}
\end{figure}

 For 26 of the events, in addition to the aggregated impact data, we also have access to data describing the damage level of individual buildings. This data is collected from both the \citet{copernicus2012} implemented by the European Commission and \citet{unosat2023}. For these events, satellite imagery is used to classify buildings within selected areas as unaffected, possibly damaged, damaged, or destroyed; with damage further classified in some cases as moderate or severe. Across the 26 events, a total of 264,109 buildings are classified, and in Figure \ref{fig:PointSummary} we show the count of the various building classifications. 

\begin{figure}[t]
\centering
\includegraphics[width=0.9\linewidth]{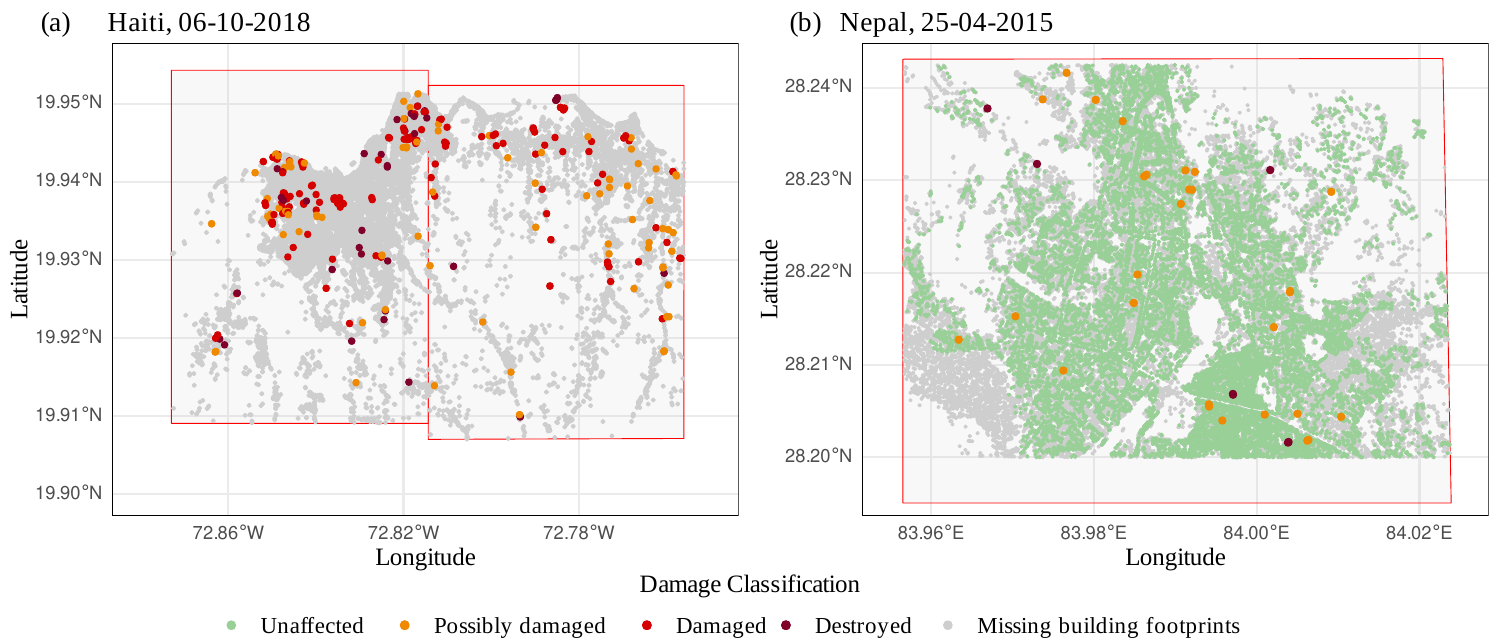}
\caption{Subfigure (a) shows the data from Copernicus for two adjacent polygons, marked in red, which were assessed for building damage following the October 2018 earthquake in Haiti. The grey points show buildings in the Bing building footprints dataset that are not included in the Copernicus data. Subfigure (b) also presents Copernicus damage data and Bing building footprints for an assessed polygon, again marked in red, now for the earthquake that occurred in Nepal in April 2015.}
\label{fig:BuildPointMAR}
\end{figure}

 The aggregated impact data often covers a broad range of hazard intensities and vulnerabilities, which can limit model identifiability, and the building point data initially presents as a promising resource to address this challenge. On closer inspection, however, there are some major issues with the building classification data across many of the events. To avoid introducing bias, the data should be missing at random, that is, the inclusion of a building in the dataset should not be dependent on the observed damage. In the extreme case, if the only buildings included are those damaged or destroyed, the model will learn to predict all buildings as damaged regardless of the hazard intensity. In Figure \ref{fig:BuildPointMAR}(a), for now ignoring the grey points, we show the building classification data issued by Copernicus for the October 2018 earthquake in Haiti. The earthquake shaking intensity in this area was below 5.75, under which we would expect relatively a small probability of damage. However, no buildings are classified as unaffected, suggesting that unaffected buildings have been omitted from the final dataset released. This issue could potentially be remedied using building footprint data, such as from Bing building footprints, to identify the missing buildings and assume that these buildings are unaffected. In the same figure, we show in grey the missing buildings identified using Bing building footprints, which amounts to around 65 thousand missing buildings. However, there are two reasons that this approach is flawed. Firstly, buildings may be missing for reasons other than the absence of damage, such as cloud cover on the satellite imagery used for classification. Assuming that these buildings are unaffected would therefore misclassify those that are damaged and missing. Secondly, Bing building footprints may identify buildings differently to Copernicus or other building footprint providers, leading to uncertainty about the number of missing buildings. In Figure \ref{fig:BuildPointMAR}(b) we display the Copernicus building classification data for the April 2015 earthquake in Nepal, which appears to be missing at random. The grey points show buildings added from Bing building footprints that do not appear in the Copernicus data, resulting in around double the number of buildings. Therefore, for this event, if we only had access to the damaged buildings and populated the missing buildings using building footprints, the probability of damage would be estimated to be around half of that when using the true missing at random data. 

\begin{figure}[t]
\centering
\includegraphics[width=0.9\linewidth]{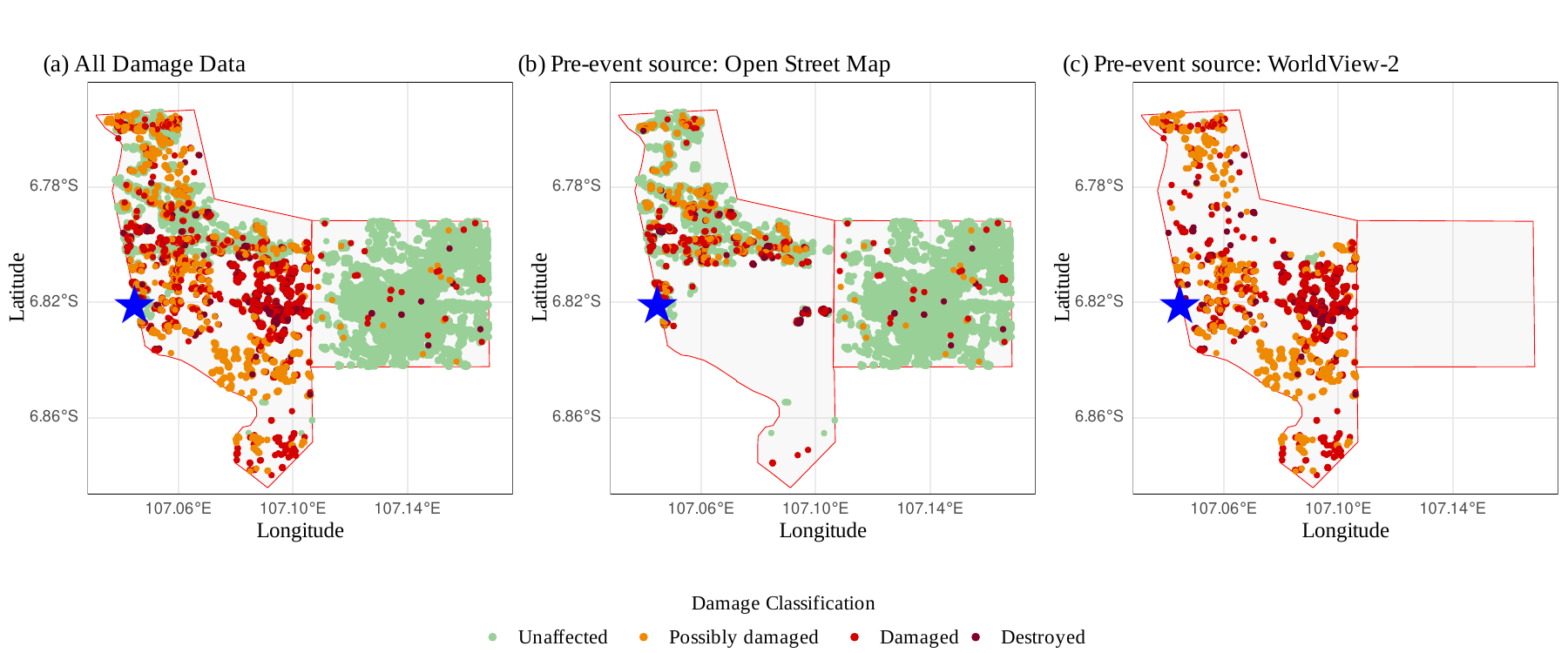}
\caption{Copernicus building damage classification data for the earthquake that occurred in West Java, Indonesia, in November of 2022. The two polygons assessed for building damage are outlined in red, and the blue star represents the location exposed to the largest shaking intensity.  Subfigure (a) shows all the damage data, subfigure (b) only shows the data with Open Street Map as the pre-event source of satellite imagery, and subfigure (c) only shows the data with Worldview-2 as the pre-event source of satellite imagery.}
\label{fig:IDN2022}
\end{figure}

An alternative solution would be to work with only the building point data that is missing at random. However, as there is no metadata providing reasons for building inclusion or missingness, and because the methodology is inconsistent even within events, deciding which datasets are missing at random can be challenging and often introduces subjective judgement. In Figure \ref{fig:IDN2022}(a) we display Copernicus building damage classification over two adjacent polygons for the earthquake that occurred in West Java, Indonesia, in November of 2022. The data initially seems as though it may be missing at random: there is a large proportion of unaffected buildings, and while the proportion is lower in some areas, these are generally closer to the epicentre. However, in Figure \ref{fig:IDN2022}(b) and \ref{fig:IDN2022}(c) we split the data based on the pre-event satellite imagery used during the damage classification. While the data using Open Street Maps appears to be missing at random, the gradings using WorldView result in a damage rate that seems unreasonably high, particularly when compared to the Open Street Maps data. However, there are still unaffected buildings in the WorldView data, making it difficult to state with confidence that the data is not missing at random. This becomes further complicated in locations where there are not various sources over which behaviour can be compared, and in higher intensity events where we expect the probability of damage to be very high regardless of missingness. 

\begin{figure}[t]
\centering
\includegraphics[width=0.7\linewidth]{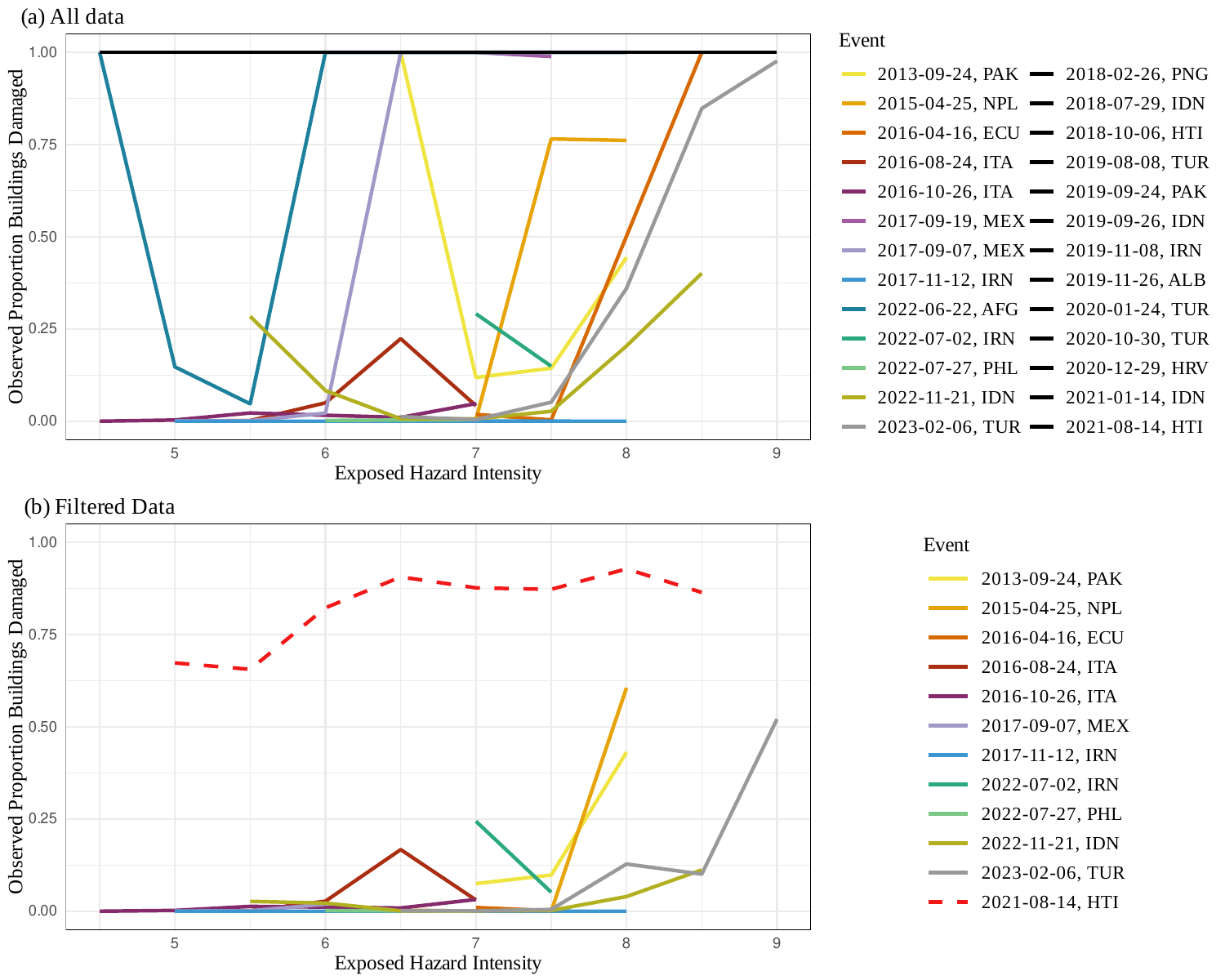}
\caption{Within each event, we group buildings by the maximum exposed hazard intensity, rounded to the nearest 0.5, and plot the observed proportion of buildings damaged against the intensity. Subfigure (a) shows all data collected across the 26 events, with events coloured in black exhibiting a damage rate of $100\%$. To produce the filtered data used in Subfigure (b), buildings have been grouped by the source and date of the satellite imagery used to detect damage, and groups with less than 10 buildings or a damage rate above $95\%$ removed. In Subfigure (b), a red dashed line has also been added to display the StEER data from the August 2021 earthquake in Haiti, which was collected using building damage data from streetview imagery and door-to-door assessment rather than satellite imagery. In both subfigures, we remove buildings identified as possibly damaged when calculating the proportion of damaged buildings.}
\label{fig:PointDatIntensity}
\end{figure}

In an endeavour to produce a dataset that is missing at random, we first group the point data within each event by all available source and date information associated with the satellite imagery. Next, we eliminate all groups containing less than 10 buildings or with a proportion of unaffected buildings less than $5\%$. In Figure \ref{fig:PointDatIntensity}, we compare the proportion of observed buildings that are damaged at different intensities under the original and filtered data. Filtering the data eliminates the erratic jumps between high and low damage proportions, and creates a dataset that appears more trustworthy. However, it has not eliminated all concerns regarding data quality. There are some events, particularly less recent ones, with limited source information, so the groupings may contain mixtures of data that are and are not missing at random. Further, it is possible that the filtering discarded data where the true proportion of unaffected buildings actually was less than $5\%$, particularly at very high intensities. In the figure, we have also included data for the Haiti earthquake in August of 2021, provided by the Structural Extreme Events Reconnaissance (StEER) Network \citep{kijewski2024}. We expect some difference between this data, collected using Streetview imagery and door-to-door assessment, to the satellite imagery, which would only reflect damage visible from above. However, there is a stark difference between the damage rates, which may reflect bias in the selection of buildings assessed for the StEER data, or instead an underreporting of damage by the satellite imagery. On account of these various concerns, we avoid using the building point data to train the model, and only use it as an alternative data form on which to evaluate model performance. 

\section{Results} 

\subsection{Results on Simulated Data} \label{sec:Testing}

\begin{figure}[t]
\centering
\includegraphics[width=0.85\linewidth]{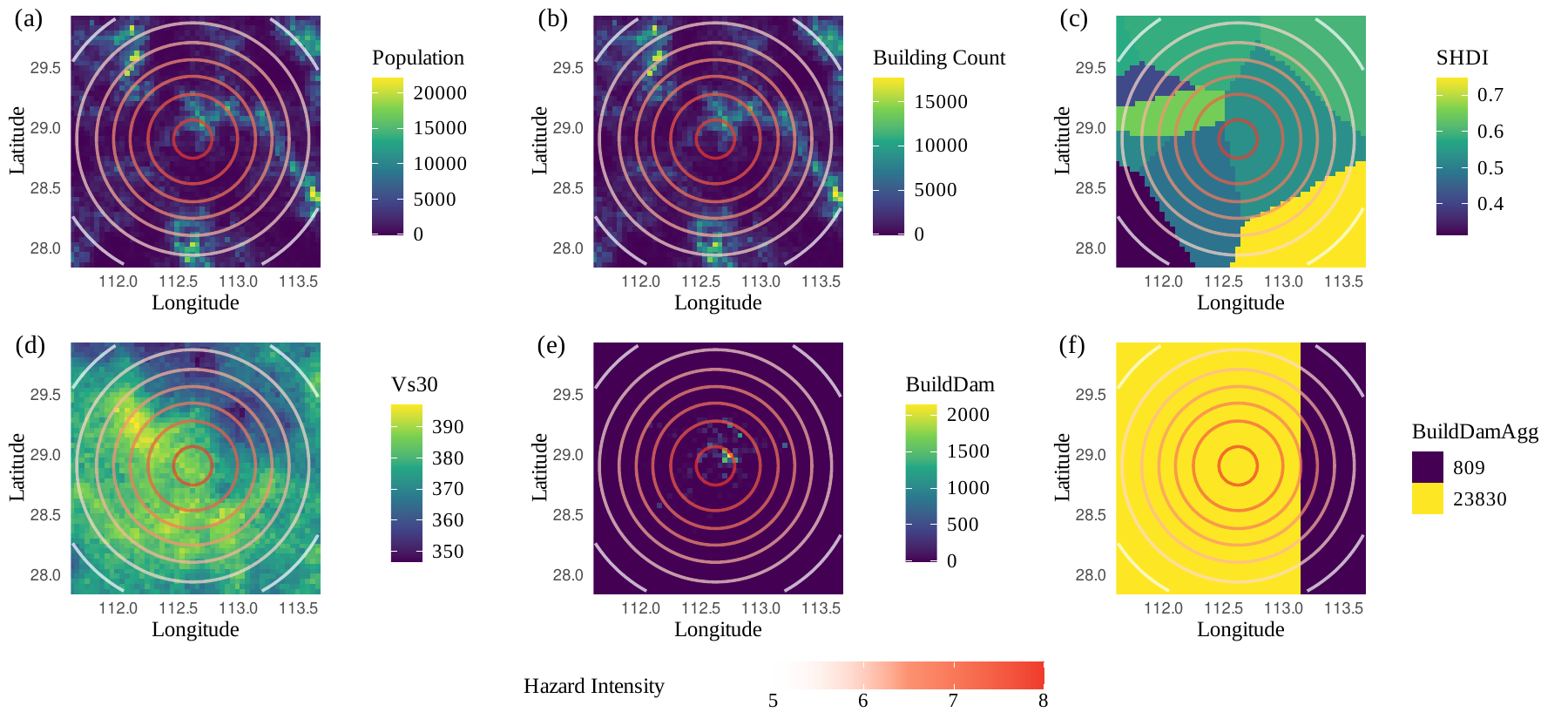}
\caption{A selection of the hazard, exposure, vulnerability, and impact data from a single simulated event. Note that while impact is sampled at the gridded resolution, it is then aggregated before fitting the model in order to mimic the real data. For example, subfigure (e) shows the sampled building damage at the gridded resolution, which has then been aggregated into two regions with total building damage counts of 809 and 23830, as shown in subfigure (f).}
\label{fig:SimEvent}
\end{figure}

To validate the methods described, we create a set of simulated events and investigate whether the algorithm is able to effectively recover the parameterisation used to generate impact data for these events. The simulated dataset is constructed to mimic properties of the dataset of 167 real earthquake events described in \citet{patten2024}, to which we will later also apply the model. For example, data for some vulnerability covariates, such as the human development index, is held constant across blocks of neighbouring grid cells, as this reflects the real-world data which is typically only available at an aggregated level. The code used to generate the hazard, exposure and vulnerability data for the simulated events is provided in the linked GitHub\footnote{\url{https://github.com/hamishwp/ODDRIN}}, and in Figure \ref{fig:SimEvent} we present an example of a single event. Note that, for computational efficiency, the maximum simulated event size is 50 by 50 grid cells, whereas the largest event in the real dataset is 225 by 316 grid cells. 

\begin{figure}[t]
\centering
\includegraphics[width=0.85\linewidth]{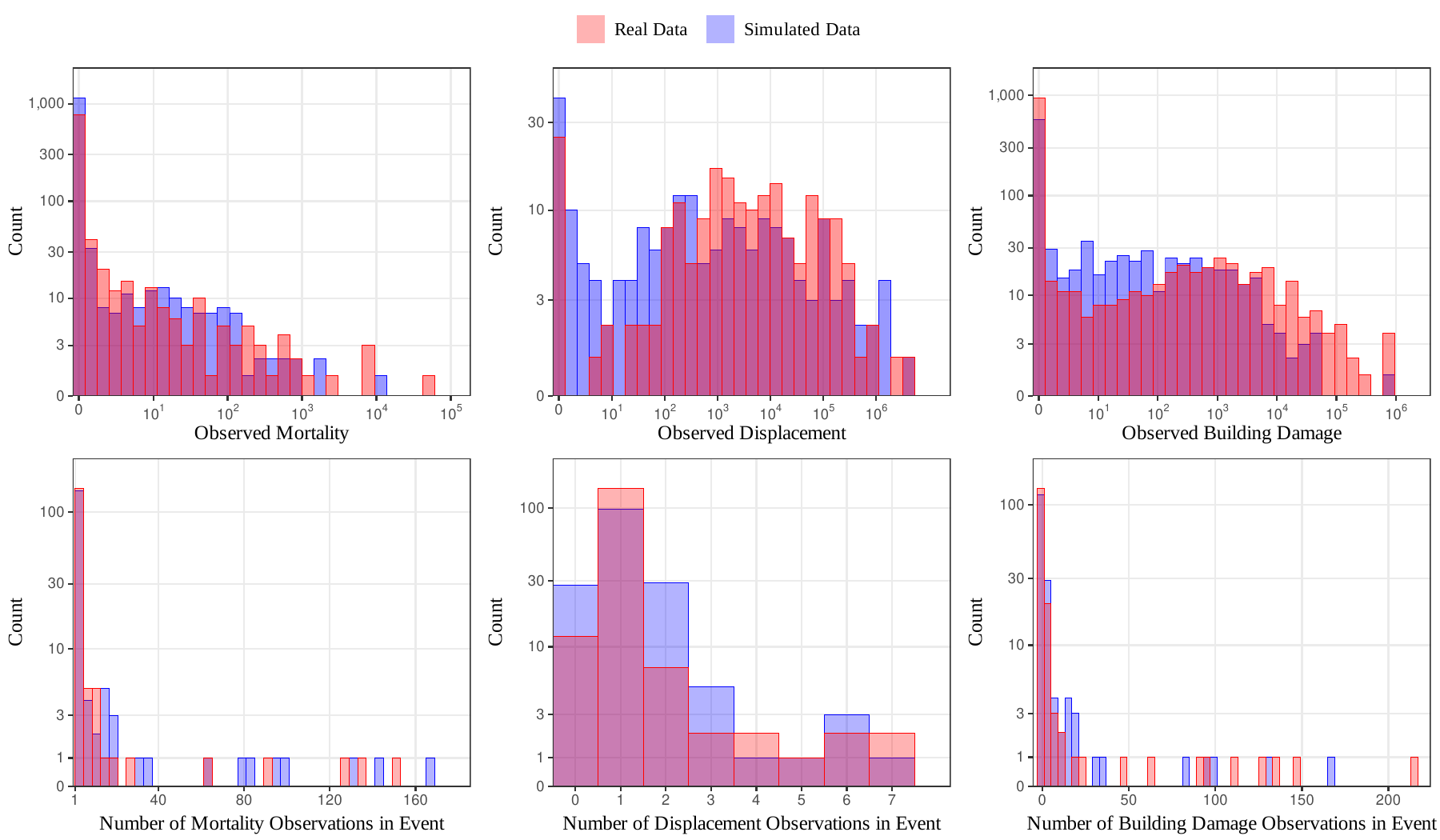}
\caption{For each of the three impact types, the histograms on the top row compare the values of all observations in the simulated data and the real data. The scaling of the x and y axes follows a pseudo-log transformation with base 10 to display zeroes together with large values, and the widths of the histogram bars correspond to equal ranges on the pseudo-log scale. The bottom row contains histograms displaying the number of observations per event. For example, the simulated data contains one event with over 160 mortality observations, reflecting a high number of subnational regions with observed impact data. The y axes again use a pseudo-log scale with base 10.}
\label{fig:SimVsObs}
\end{figure}

To produce impact data for each simulated event, we sample from the model using a fixed parameterisation. The impact data is then aggregated across grid cells, sometimes completely to produce a total impact observation, and sometimes in small clusters to mimic subnational impact data. In Figure \ref{fig:SimVsObs}, we compare the impact data from the 167 simulated events to the dataset of 167 past earthquake events. There are some differences between the two datasets, for example, an increased proportion of zeros in the observed building damage for the real data. Overall, however, the simulated dataset is sufficiently similar to the real dataset that we expect it would be a reasonable test for the performance of the proposed method. When fitting the model, we train the model on two thirds of the dataset (112 events) and hold-out the remaining one third (67 events) for testing. 

\begin{figure}[t]
\centering
\includegraphics[width=0.55\linewidth]{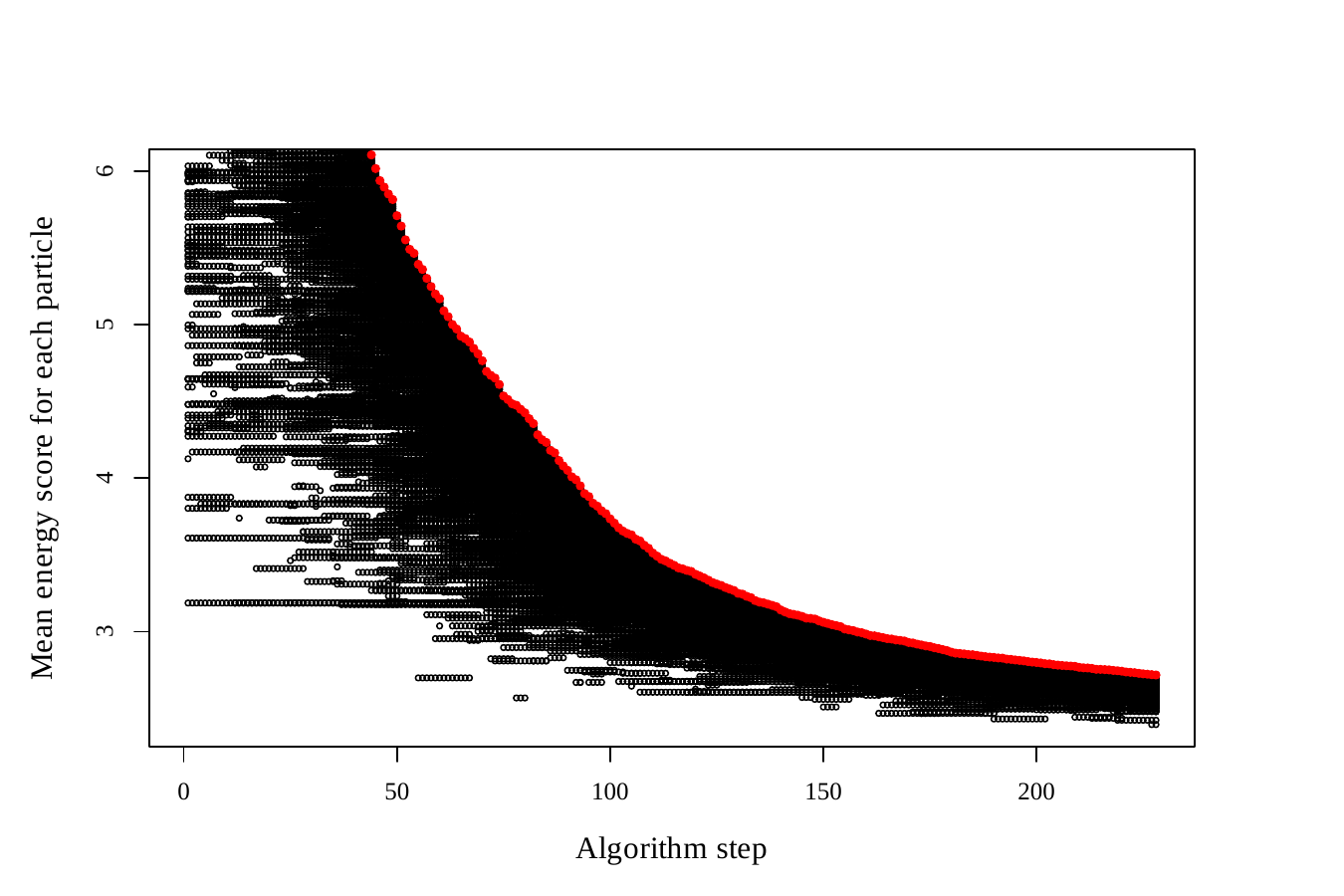}
\caption{For each step in the ABC-SMC algorithm, we display the mean energy score for each of the 1000 particles. The red points show the tolerance $\delta$ at each step. The tolerance sequence begins above 50, but we crop the y-axis to better display behaviour in the latter steps.}
\label{fig:DvsStep}
\end{figure}

\begin{figure}[t]
\centering
\includegraphics[width=0.75\linewidth]{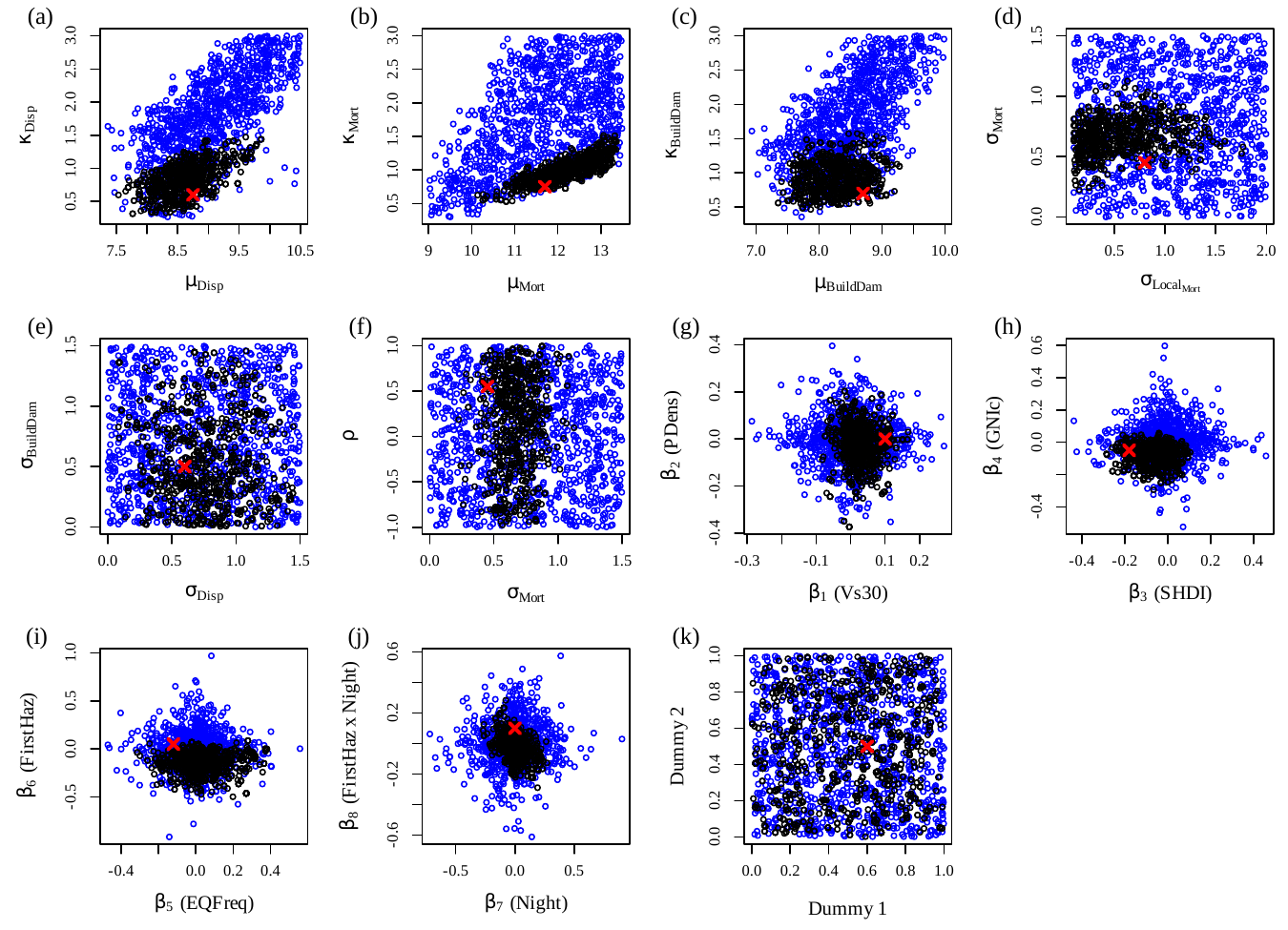}
\caption{A comparison between samples from the prior distribution (blue), posterior distribution (black) and the parameter value used to generate the data (red) for the simulated events. The parameters are paired to show correlation in places, but note that parameters are often also correlated outside of these pairings.}
\label{fig:SimPosteriors}
\end{figure}

\begin{figure}[t]
\centering
\includegraphics[width=0.7\linewidth]{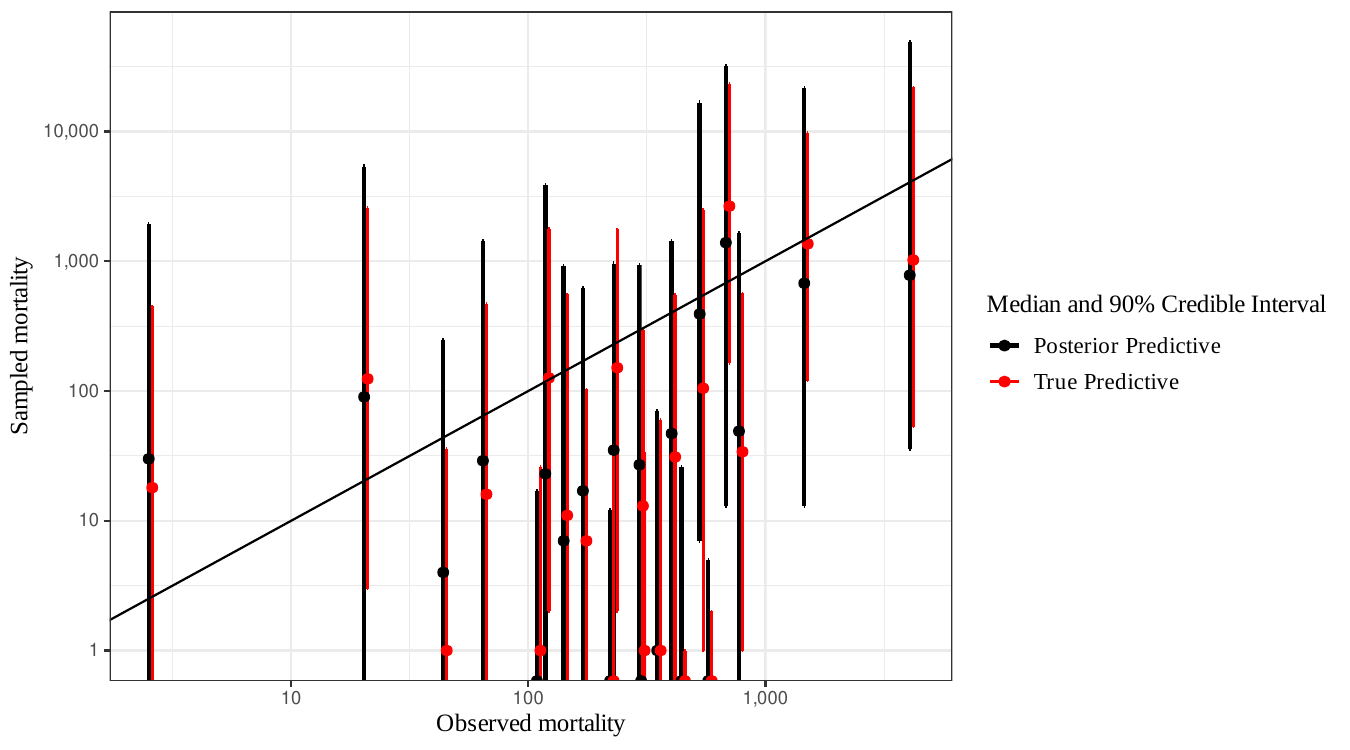}
\caption{A comparison between the posterior predictive distribution (black) and the predictive distributions under the true parameter values (red) for the set of testing events. The points display the median prediction and the bars display $90\%$ intervals. We omit events with observations of $0$ as these are tightly clustered and challenging to differentiate.}
\label{fig:PostPredSim}
\end{figure}

We ran the algorithm for $228$ steps, by which point the decrease in tolerance was less than $0.1\%$ between steps, and used the particles at the 228\textsuperscript{th} step to form the approximate posterior distribution. In Figure \ref{fig:DvsStep} we display the mean energy score for each particle as the algorithm progresses. This demonstrates an effective traversal of the parameter space, with the mean energy scores evaluated for particles at the final step all less than those calculated at the first step, and in many cases significantly so. In Figure \ref{fig:SimPosteriors} we compare the prior and posterior distributions for each parameter. We also compare prior and posterior samples for two dummy variables that are not included in the model, and confirm that the posterior distribution remains similar to the prior. For all parameters the posterior samples do contain the true parameter values; however, in some cases the posterior remains diffuse, or the true parameter lies near the boundary. There are a few reasons for this. First, the large variance of the local and event-wide error terms, while reflecting variability of the real data, also makes the parameter values hard to identify. Second, one weakness of the energy score is that it is poor at detecting incorrectly specified correlations in joint distributions, leading to difficulties identifying the correlation term $\rho$ used in the event-wide error covariance matrix. Thirdly, there is correlation between many of the parameters, making parameter identifiability challenging. For example, since the impact in each grid cell must be non-negative, increasing the error variances or vulnerability covariates works asymmetrically and shifts the impact distribution upwards, which can be counteracted by shifting the location of the damage curves. In Figure \ref{fig:PostPredSim}, we compare posterior predictive distributions over the held-out testing set of events to predictive distributions under the true parameter values. In general, the true predictive distributions are smaller than the posterior predictive ones. This is unsurprising as the posterior predictive distributions will also convey parameter uncertainty. However, on the whole, the distributions are similar, demonstrating that even if identifiability challenges limit inference on some parameters, the predictive performance of the fitted model is still relatively strong. 

\subsection{Results on Past Earthquake Events} \label{sec:Results}

The model is trained on 112 events and tested on the remaining 67. To ensure a mix of impact sizes within the training and testing sets, we sort the events by total mortality and allocate every third event to the testing set. The ABC-SMC algorithm is run using the same hyperparameters as for the simulated data, but now for 170 steps. In Appendix \ref{appendix:mcmc_comparison} we compare the posterior distributions to those obtained using an MCMC algorithm targeting the loss function without the threshold adaption,
\begin{equation}
p_{\mathrm{LF}}(\theta, x  \mid y) \propto \exp \left(-\omega\frac{1}{N} \sum_{n=1}^N \mathrm{SR}(y_n, x_{n,1:M}) \right)  p(x \mid \theta) p(\theta). \label{eq:scoringrulepost_app}
\end{equation}
The two approaches yield similar posterior distributions, supporting the validity of the ABC-SMC algorithm. In Appendix \ref{appendix:mcmc_comparison} we discuss the MCMC algorithm used in more detail, and also provide trace plots to explore the mixing of the MCMC chains. We also discuss the computational demand of the two algorithms; while the ABC-MCMC algorithm requires fewer samples from the likelihood in total, the ABC-SMC algorithm provides greater scope for parallelisation that can reduce time. For the remainder of this section, we refer to the results of the ABC-SMC algorithm. 

\begin{figure}[t]
\centering
\includegraphics[width=0.9\linewidth]{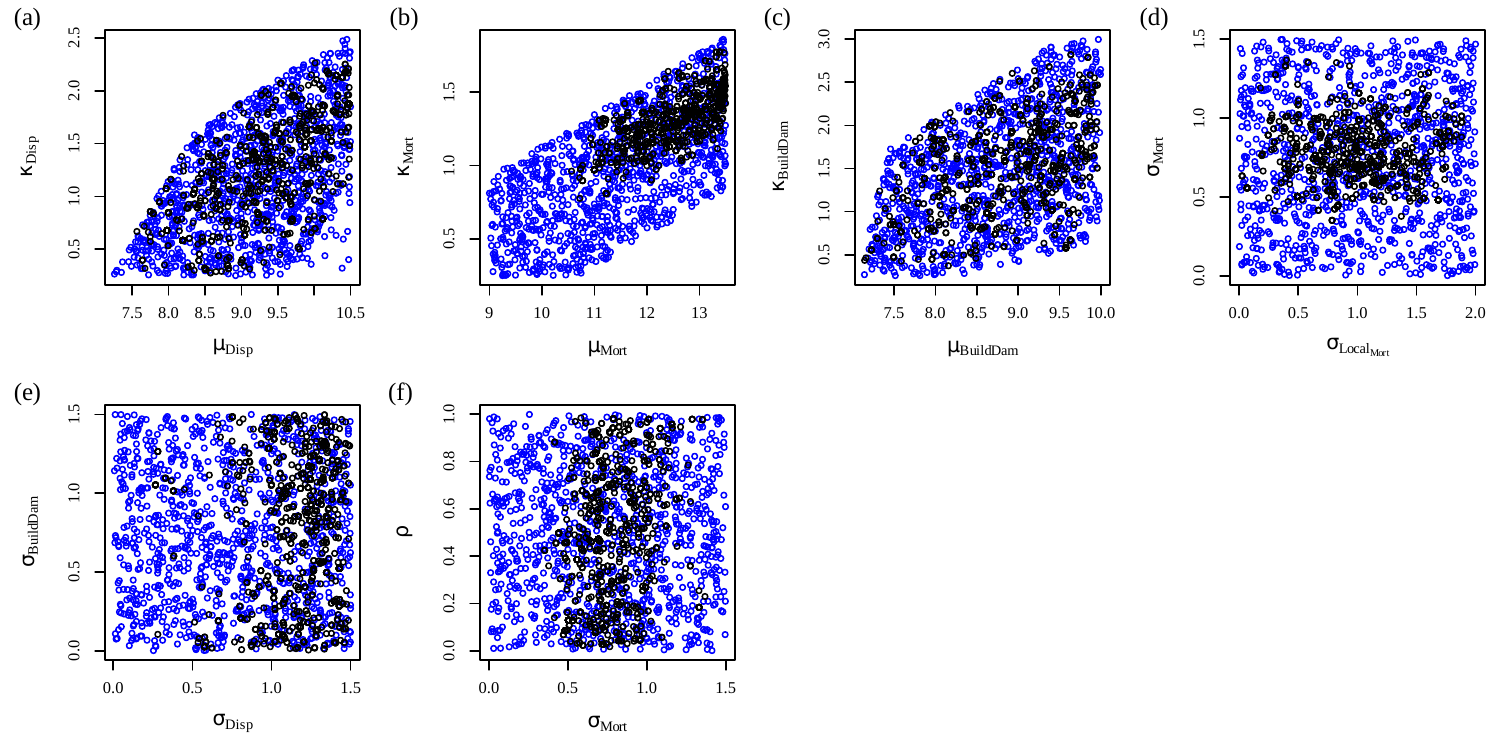}
\caption{A comparison between samples from the prior distribution (blue) and posterior distribution (black) for the model trained on real impact data from 112 past events. Posterior distributions for the vulnerability covariate coefficients are shown separately in Figure \ref{fig:VulnPosteriors}.}
\label{fig:RealPosteriors}
\end{figure}

\begin{figure}[t]
\centering
\includegraphics[width=0.9\linewidth]{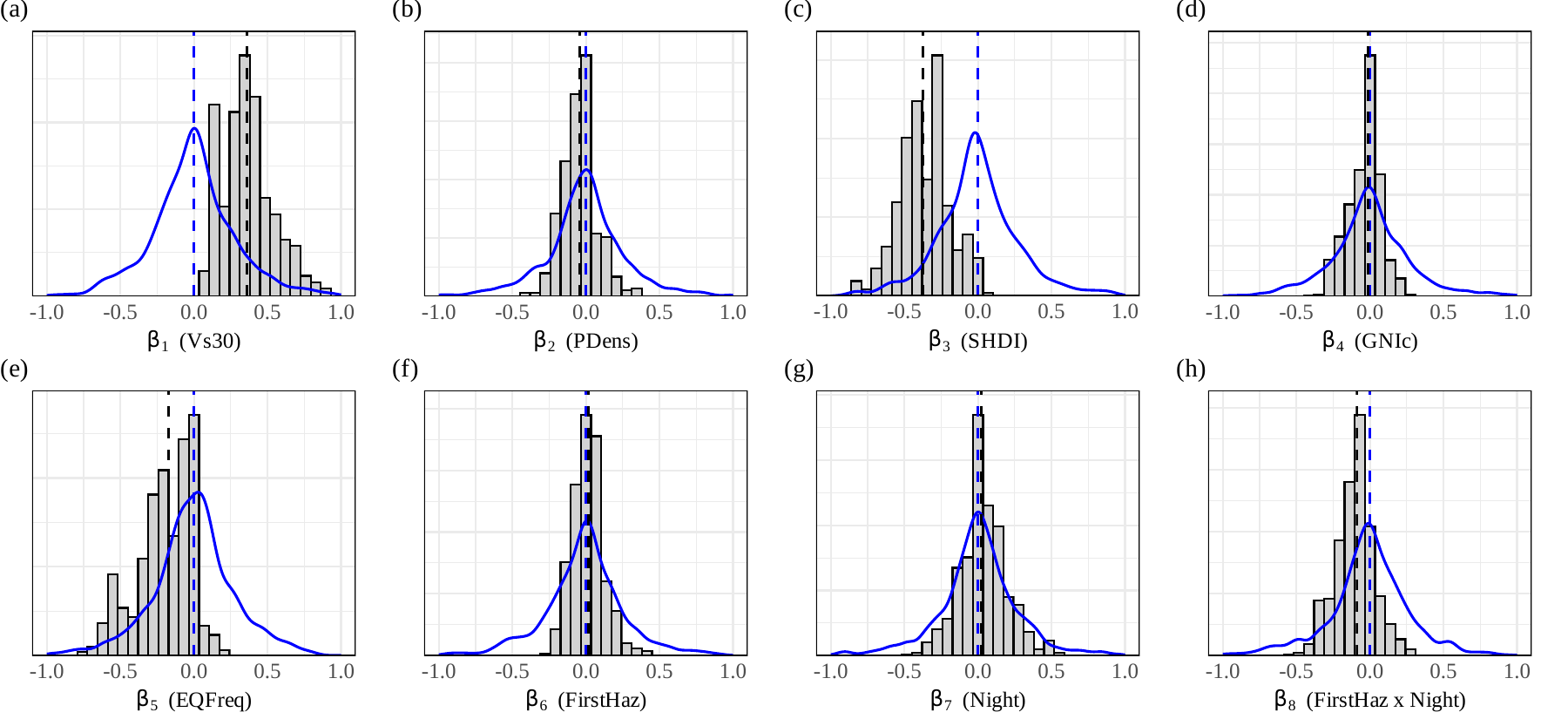}
\caption{For each of the vulnerability covariate coefficients, we compare the prior distribution, shown using a blue line, and the posterior distribution, shown using a grey histogram. The black dashed line shows the median of the posterior distribution. The prior distributions are centred over zero, which is also marked using a red dashed line. While many posterior distributions remain close to the prior, others shift more above or below zero, particularly those for $\beta_1$ (Vs30) and $\beta_3$ (SHDI), indicating a relationship between the relevant covariate and disaster impact. 
}
\label{fig:VulnPosteriors}
\end{figure}

We compare the prior and posterior distributions in Figures \ref{fig:RealPosteriors} and \ref{fig:VulnPosteriors}, with the latter focussing on the vulnerability covariate coefficients. The posterior distributions for the parameters related to mortality are far more concentrated than those for displacement and building damage; this is due to the noisiness of the displacement and building damage data and their reduced weighting in calculating the energy score. Some parameters, such as $\mu_{\text{Mort}}$, $\sigma_{\text{Disp}}$ and $\sigma_{\text{BuildDam}}$ have reasonable amounts of posterior weight near their upper bound. Given the careful construction of the prior distributions, it is more likely that this reflects identifiability challenges or an adjustment for measurement error, rather than misspecified priors. 

Regarding the vulnerability covariates, the posterior distribution for \verb|Vs30| exhibits the largest shift from the prior, with 100\% of the posterior weight lying above zero. While high \verb|Vs30| is typically associated with lower shaking intensity, this is already accounted for in the ShakeMap shaking estimation. The result may therefore reflect an overcorrection by ShakeMap, or that \verb|Vs30| is being used as a proxy for another factor, such as elevation or the mandating of better building codes in low \verb|Vs30| locations. The posteriors for \verb|SHDI| and \verb|EQFreq| also exhibit large shifts from the priors, with 1.1\% and 7.7\% of the posterior weight lying above zero respectively. Comparing the prior and posterior evidence that the coefficients are positive gives respective Bayes factors of 0.01 and 0.08. This provides strong evidence for a negative relationship between these variables and damage. These results align with intuitive reasoning; we would expect the impact to be larger in disadvantaged areas where earthquakes occur less frequently. The posterior distributions for the remaining parameters (\verb|PDens|, \verb|GNIc|, \verb|FirstHaz|, \verb|Night| and \verb|FirstHaz|) remain largely centred around zero. Surprisingly, the posterior for the interaction term \verb|FirstHaz| $\times$ \verb|Night| has shifted slightly below zero. This shift is unexpected and likely due to a spurious correlation in the training data.

There are some parameters for which the posterior remains relatively unchanged from the prior, such as the correlation term $\rho$ between impact error types, and the coefficient for the population density term, $\beta_2$. While it may be tempting to remove these parameters from the model when sampling from the posterior predictive distribution, there are two reasons that this should be avoided. Firstly, fixing these parameters, such as setting $\rho=0.5$, would remove uncertainty about the parameters that should be present within our predictions. Secondly, on account of the identifiability challenges and associated parameter correlation, removing one parameter without adjusting another may affect the model fit. Before future model fits, particularly if the model becomes more complex, it may be worth removing the vulnerability covariates with Bayes factors close to one. 

\begin{figure}[t]
\centering
\includegraphics[width=\linewidth]{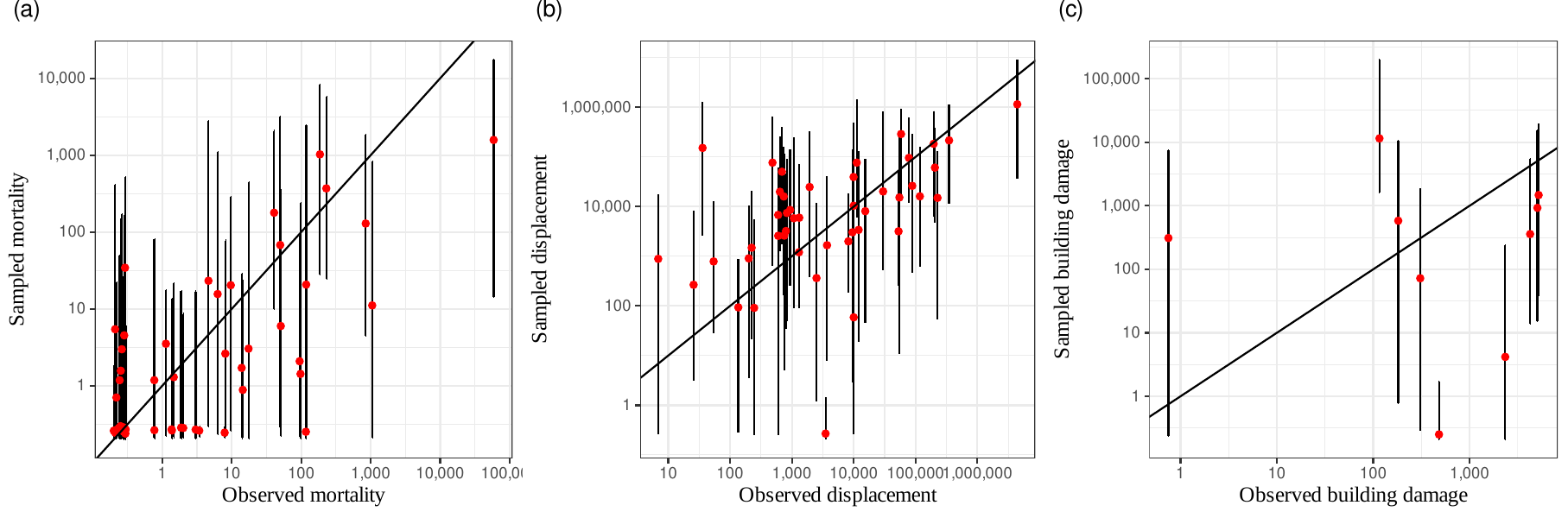}
\caption{A comparison between the observed total impact and samples from the posterior predictive destributions for the testing events. The red points show the median of the posterior predictive distribution and the bars show the 90\% credible interval.}
\label{fig:PostPredReal}
\end{figure}

In Figure \ref{fig:PostPredReal} we display the medians and 90\% credible intervals of the posterior predictive distributions for the total aggregated impact across the testing events. Prediction for mortality performs relatively well, with predictions generally increasing with the observations and about $85\%$ of the $90\%$ credible intervals containing the observed value. This suggests a slight underdispersion, with some observations lying at the very upper end of the posterior predictive distributions. The predictive distributions for displacement also perform fairly well and are similarly calibrated, with $84\%$ of the $90\%$ credible intervals containing the observed values. The predictive distributions over the building damage perform poorest, with no clear trend between observed and predicted damage and three clearly outlying observations amongst a relatively small testing set. It is challenging to assess how much of this error should be attributed to measurement error, uncertainty in the initial building count, or weaknesses in the model. The credible intervals across all impact types are relatively large, typically covering many orders of magnitude. This uncertainty can be attributed to numerous sources, including uncertainty in the available data (such as population count, hazard intensity and building count), the absence of data describing covariates that may be relevant, and inherent randomness, also referred to as aleatoric uncertainty, that can be modelled but not reduced. 

We compare the predictive performance of ODDRIN to two leading impact estimation tools commonly used in the field, PAGER and GDACS. As neither GDACS nor PAGER predict displacement or building damage (outside of economic value), we compare all three models on their prediction of mortality. Note that it is challenging to construct a fair comparison of the three models, as we do not have access to the original predictions issued by those models; some of the GDACS and PAGER predictions were updated as new information about the estimated disaster impact became available, information that is not available to ODDRIN. 
GDACS issues an alert of green for predicted fatalities between 0 and 10, orange for 10 to 100, and red for over 100. To perform comparison, we apply the same labelling system to the median posterior predicted mortality from ODDRIN. Over the 50 events for which GDACS issues predictions, ODDRIN makes correct predictions for $80\%$ of the events, and GDACS for $67\%$. PAGER issues more detailed mortality predictions, providing a discrete predictive distribution in which  probabilities are assigned to fatality counts in the intervals $[0,1)$, $[1, 10)$, $[10, 10^2)$, $[10^3,10^4)$, $[10^4, 10^5)$, and  $[10^5,10^7)$. PAGER mortality predictions do not include fatalities caused by landslides; however, adjusting the mortality by the number of landslide fatalities using data from \citet{seal2022comprehensive} does not change the interval for any of the testing data points. We discretise the ODDRIN posterior predictive distributions onto the same intervals and evaluate both ODDRIN and PAGER using the ranked probability score (RPS), a scoring rule used to compare an outcome with a set of ordinal categories with assigned probabilities. Over the $53$ events for which PAGER issued predictions, we find a similar performance between the two, with a mean RPS for ODDRIN of 0.327 and a mean RPS for PAGER of 0.365 (lower scores being better). This demonstrates a similar or slightly improved performance of ODDRIN compared to a leading impact estimation tool in the prediction of total mortality. 


\begin{figure}[t]
\centering
\includegraphics[width=0.9\linewidth]{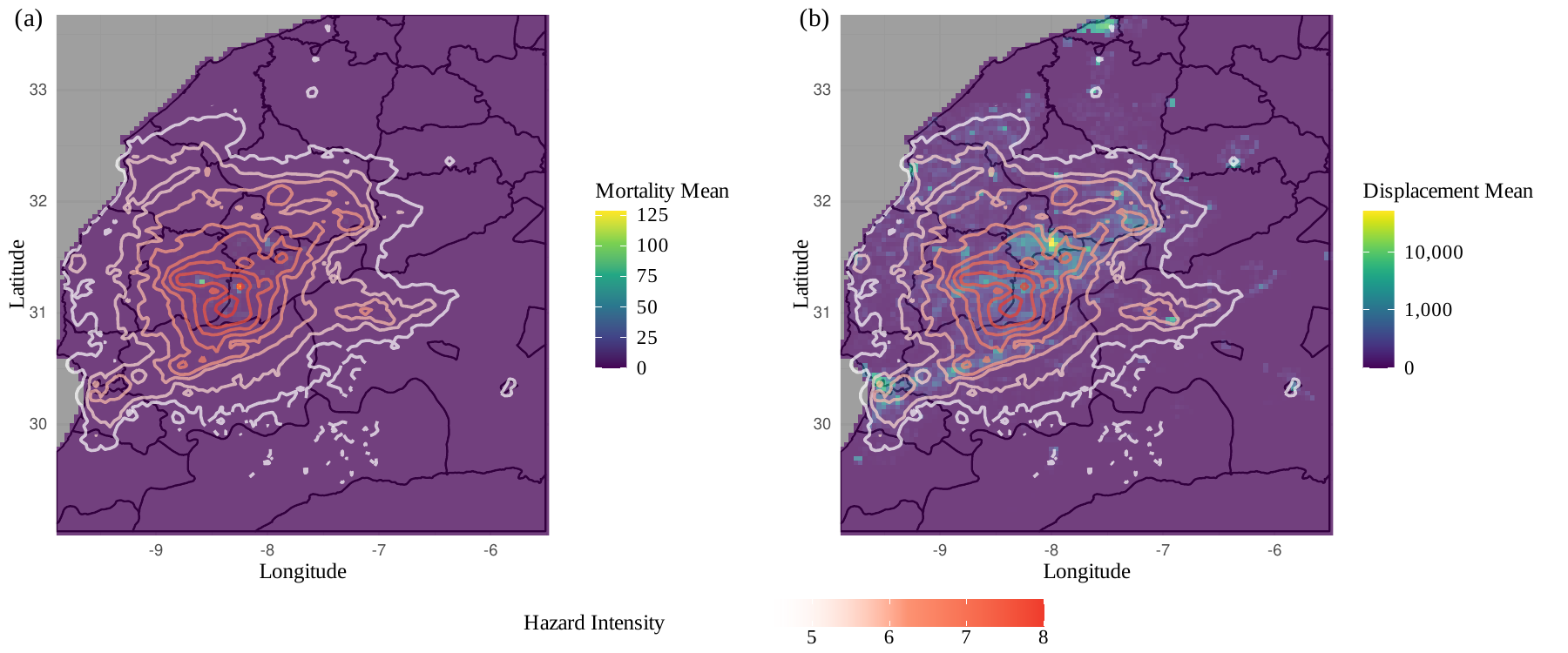}
\caption{Means of the posterior predictive distributions for mortality and displacement in each exposed 2.5 by 2.5 arcminute grid cell for the earthquake in Morocco on September 8\textsuperscript{th}, 2023.}
\label{fig:MoroccoMort}
\end{figure}

\begin{table}
\centering
\begin{tabular}{ l r r r r r}
 \hline
\textbf{Region} & \textbf{Observed} & \textbf{Post. Median} & \textbf{Post. $5\%$ Quantile} & \textbf{Post. $95\%$ Quantile} \\
Agadir-Idaou Tanane, Souss-Massa-Draâ & 0 & 0 & 0 & 10 \\
Al Haouz, Marrakech-Tensift-AlHaouz & 1684 & 124 & 2 & 2200 \\
Azilal, Tadla-Azilal & 11 & 0 & 0 & 4 \\
Béni Mellal, Tadla-Azilal & 0 & 0 & 0 & 1 \\
Casablanca, Grand Casablanca & 0 & 0 & 0 & 3 \\
Chichaoua, Marrakech-Tensift-AlHaouz & 202 & 94 & 3 & 2350 \\
Chtouka-AïtBaha, Souss-Massa-Draâ & 0 & 0 & 0 & 4 \\
ElJadida, Doukkala-Abda & 0 & 0 & 0 & 2 \\
El Kelaâdes Sraghna, Marrakech-Tensift-AlHaouz & 0 & 1 & 0 & 35 \\
Essaouira, Marrakech-Tensift-AlHaouz & 0 & 0 & 0 & 13 \\
Inezgane-AïtMelloul, Souss-Massa-Draâ & 0 & 0 & 0 & 13 \\
Marrakech, Marrakech-Tensift-AlHaouz & 15 & 6 & 0 & 276 \\
Ouarzazate, Souss-Massa-Draâ & 41 & 0 & 0 & 13 \\
Safi, Doukkala-Abda & 0 & 0 & 0 & 27 \\
Settat, Chaouia-Ouardigha & 0 & 0 & 0 & 1 \\
Taroudant, Souss-Massa-Draâ & 980 & 18 & 0 & 435 \\
Tata, Guelmim-Es-Semara & 0 & 0 & 0 & 2 \\
Tiznit, Souss-Massa-Draâ & 0 & 0 & 0 & 1 \\
Zagora, Souss-Massa-Draâ & 0 & 0 & 0 & 1 \\
Total & 2946 & 228 & 5 & 4235 \\
 \hline
\end{tabular}
 \caption{The observed mortality in each exposed administrative level two region, as well as the medians, $2.5\%$ quantiles, and $97.5\%$ quantiles of the posterior predictive distributions for the mortality. We also include the total mortality. Most of the $95\%$ credible intervals contain the observed value, except for Taroudant, where the mortality is underestimated. }
 \label{table:SubNatQuantiles}
\end{table}

In addition to total impact estimates, as issued by GDACS and PAGER, ODDRIN also seeks to support decision-making during response coordination via the provision of full spatial impact estimates. In Figure \ref{fig:MoroccoMort} we display the mean of the posterior predictive prediction for mortality issued for the earthquake that occurred in Morocco on September 8\textsuperscript{th}, 2023. We use the mean here, rather than the median as previously, as the median prediction for mortality is zero in most locations except near the epicentre. 
Validation of the gridded predictions is challenging as impact data is rarely available at this resolution. However, when aggregating using the subnational level two administrative boundaries, we find that the $90\%$ posterior predictive credible intervals contain the observations for 17 of the 19 subnational regions ($89\%$) exposed to an intensity larger than 4.5. Table \ref{table:SubNatQuantiles} presents the observed mortality in each region, as well as the median, $5\%$ quantiles and $95\%$ quantiles of the posterior predictive distribution.  

\subsection{Results on Point Building Data} \label{sec:PointResults}

\begin{figure}[t]
\centering
\includegraphics[width=0.7\linewidth]{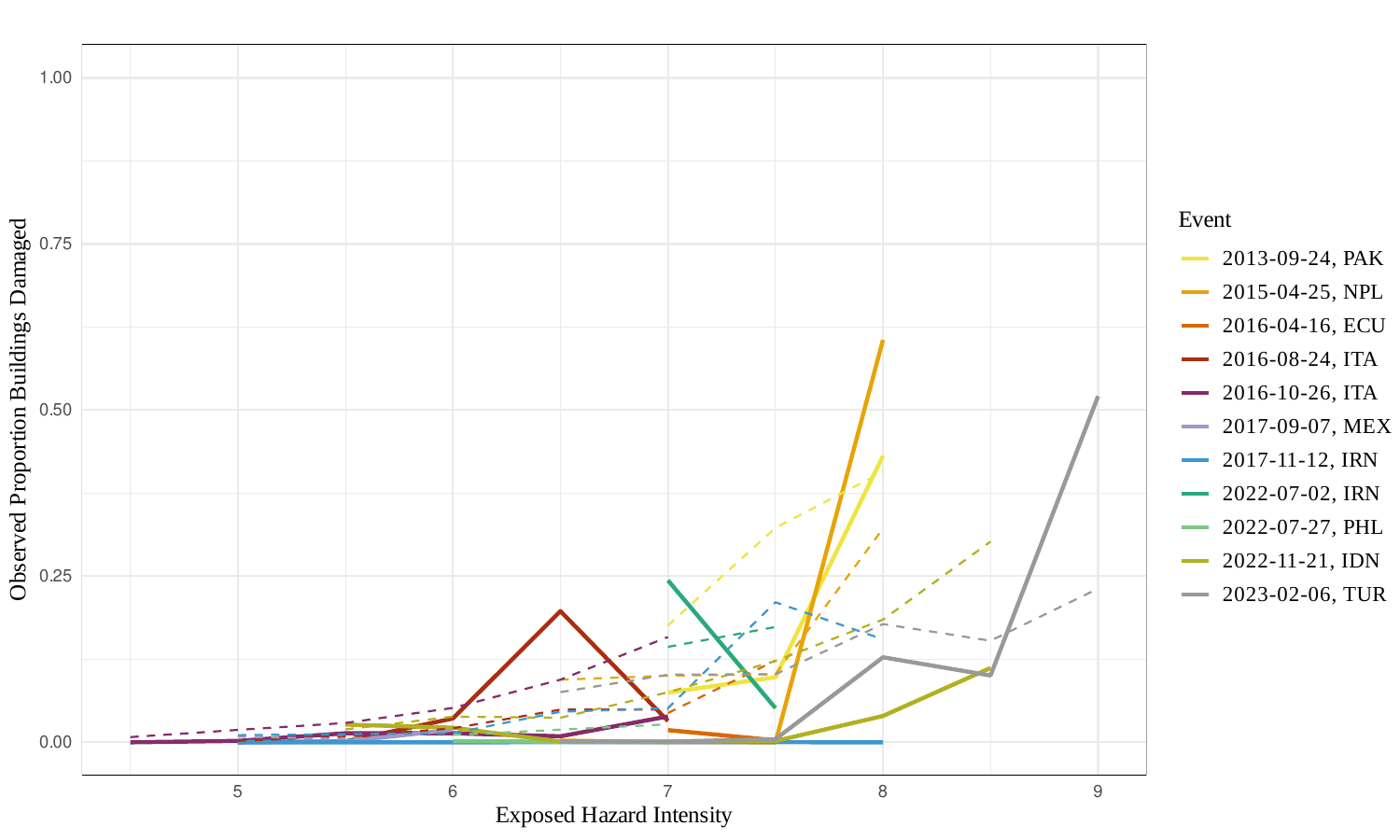}
\caption{To produce the solid lines, buildings are grouped within each event by their maximum exposed intensity, rounded to the nearest 0.5, and the proportion of damaged buildings calculated. The dashed lines show the average of the median posterior probability in each group.}
\label{fig:PostMedianProbs}
\end{figure}

We also evaluate the model performance on the point building data introduced in Section \ref{sec:AggImpactData}. We use the same model as for aggregated building damage to find the probability of building damage for buildings within each grid cell. Given that the model does not account for building characteristics such as the building materials, age, or number of storeys, we would not expect it to perform well in identifying damage in one building compared to adjacent buildings. However, we can compare the observed proportion of buildings damaged within an area to the modelled probability. To achieve this, we combine buildings within the 2.5 $\times$ 2.5 arcminute grid cells used to perform modelling, and take the posterior median of the damage probability within each grid cell. To ease analysis, we also remove buildings from the dataset marked as possibly damaged. In Figure \ref{fig:PostMedianProbs} we group grid cells by event and exposed intensity, rounding to the nearest 0.5, and plot the observed proportion of buildings damaged against the exposed intensity. We also display the median posterior damage probabilities, which have been grouped in the same way but averaged over all buildings in the group. The probabilities appear to provide a reasonable fit, however, the posterior probability of damage increases faster at a lower hazard intensity than many of the observed damage proportions. This may reflect that training the model on the aggregated data has produced a weak fit, perhaps due to poor data quality or issues with identifiability when cumulating the impact over a range of exposed intensities. Conversely, it may reflect that the aerial-view imagery is not showing damage that can be observed on the ground, and is thus under-reporting the true damage. The latter aligns with the findings of \citet{kerle2010}, which showed that observations of damage from vertical and ground-level perspectives diverge at lower damage levels. 

\begin{figure}[t]
\centering
\includegraphics[width=0.5\linewidth]{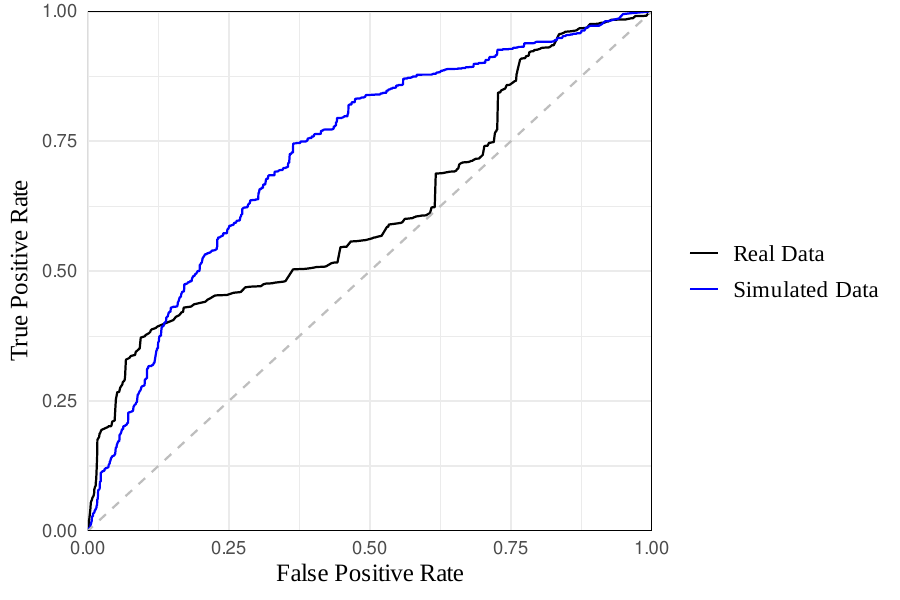}
\caption{A comparison between the ROC curves for the real data and the simulated data. The grey diagonal line demonstrates the expected performance of a random classifier.}
\label{fig:ROCcurves}
\end{figure}

We also produce a receiver-operating characteristic (ROC) curve which compares the true positive rate and false positive rate of damage identification performed by the model. A ROC curve is produced by varying the threshold required for positive identification between 0 and 1. In this case, if the modelled probability of a building experiencing damage, $p^{\text{BuildDam}}$, is larger than the threshold then the building is classified as damaged. Given the variability of $p^{\text{BuildDam}}$, we use the median values over 50 posterior samples. In Figure \ref{fig:ROCcurves} we display the ROC curves produced when classifying buildings in the real and simulated point building datasets. For the simulated dataset we calculate $p^{\text{BuildDam}}$ using the true parameter values used to generate the data. Even under the true parameter values, $p^{\text{BuildDam}}$ is variable due to the local and event-wide error terms, so we again take the median over 50 samples. Note that for the simulated data, the error terms are the only source of misidentification, so the ROC curve could be improved by reducing their variance when generating the data. However, the current values provide a reasonable benchmark on which to compare the real data ROC curve, which demonstrates a fairly strong performance at identifying the buildings that are most likely to be either unaffected or damaged, but weaker performance where it is less clear. The area under the ROC curve (AUC) values for the real and simulated data are 0.62 and 0.72 respectively. 

\section{Conclusion}\label{sec:Conclusion}

We have developed a Bayesian approach to earthquake impact modelling, adapting an existing scoring-rule based loss function to implement an ABC-SMC algorithm in a setting where the use of summary statistics would result in significant information loss. The methods were applied to a set of simulated events, and despite inference being weaker over some parameters, the fitted model generally demonstrated strong predictive capabilities. 
We also fitted the model to a set of real past earthquakes ranging from 2011 to 2023 and spanning 76 countries. While predictive performance for building damage was weak, performance appears relatively strong and well-calibrated for mortality and displacement. Comparing to two leading impact estimation tools, GDACS and PAGER, we found improved performance in the prediction of total mortality over a set of held-out events relative to GDACS and comparable performance to PAGER. Note that this is a conservative benchmark, as some GDACS and PAGER predictions are updated when information about the disaster become available. ODDRIN poses a number of further benefits over these existing tools, including the gridded spatial mapping of impact predictions within an empirical framework, robust uncertainty quantification within a Bayesian framework, the capacity to handle events with foreshocks and aftershocks, and the provision of a joint model across impact types. 

ODDRIN presents a strong foundation for an open-source, statistically validated disaster impact estimation tool. The methodology is flexible and, following appropriate model adjustments, could be applied to a range of other hazards such as heatwaves, floods, and tropical cyclones. Focussing on earthquakes, there are also numerous avenues for further investigation that may improve ODDRIN's performance. The model currently assumes independence of the error terms between grid cells. The sources of data uncertainty make this assumption unlikely; for example, there is likely to be correlation between adjacent grid cells in the errors of the ShakeMap or exposure data, as well as in vulnerability factors that are not captured via the modelled covariates. These challenges could be addressed via the availability and incorporation of data uncertainty information, or through the introduction of correlation between the grid-level error terms, for example via a Gaussian process. There are also other data sources that could be integrated into ODDRIN. The use of more impact data, for example using the newly developed Montandon - Global Crisis Data Bank database, hosted by the International Federation of Red Cross Red Crescent Societies (IFRC), may provide better quality displacement and building damage impact data, and also provide more events on which to train the model. In particular, the use of better quality building damage data would improve predictive performance for this impact type. The inclusion of data related to vulnerability covariates that more directly describe building standards or materials would also support prediction. Finally, the underlying model is flexible, and could be extended or refined in numerous directions, such as the use of non-parametric curves to translate the latent damage term to impact probabilities, exploring the use of different distributions over the error terms, or model averaging over ODDRIN and alternative machine learning models. 

\bibliographystyle{abbrvnat}
\bibliography{oup-authoring-template}


\begin{appendices}

\section{Bias of energy score calculation with few samples}\label{appendix:es_bias}

\begin{figure}[t]
\centering
\includegraphics[width=0.6\linewidth]{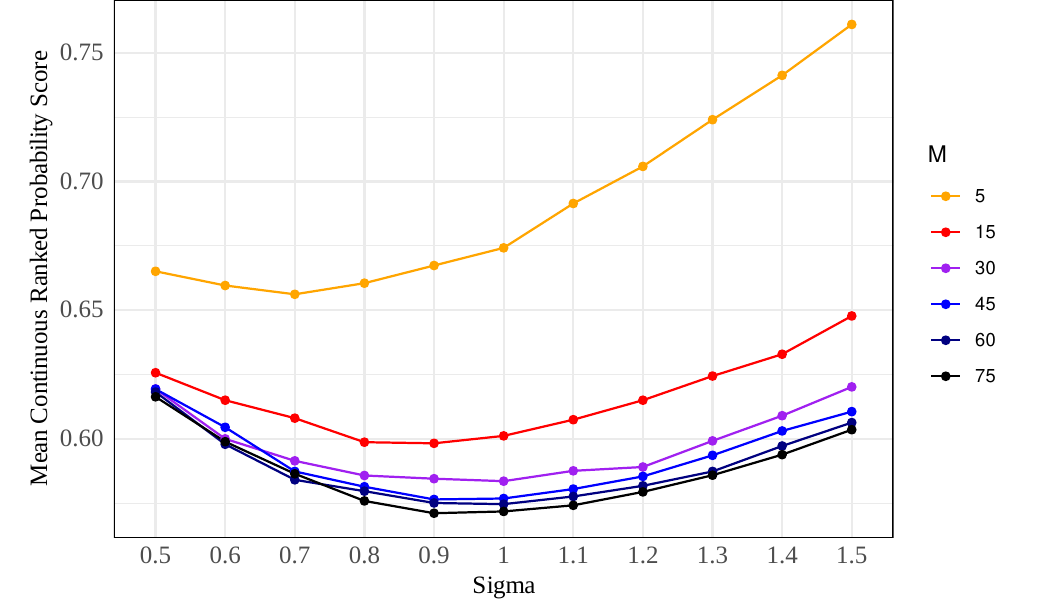}
\caption{The mean continuous ranked probability score across 100,000 trials, where the true observations are generated with standard deviation $1$ and the samples are of size $M$ and have standard deviation $\text{Sigma}$.}
\label{fig:crps_bias}
\end{figure}
When evaluating the energy score with few samples $M$ from the predictive distribution, we have found that the score is optimised at under-dispersed distributions. In Figure \ref{fig:crps_bias}, we generate $100,000$ observations from a normal distribution with mean $0$ and standard deviation $1$. For each observation, we generate $M$ samples from a normal distribution with mean $0$ and standard deviation $\sigma$, and calculate the continuous ranked probability score (the univariate instance of the energy score) comparing the samples to the observation. The plot in the figure represents the mean score over the $100,000$ observations, and shows that for $M$ less than around $30$ the scoring rule rewards $\sigma < 1$. Similar results are found with other distributions, including skewed distributions such as the log-normal distribution, which are likely to be more similar to the data generating process in our context. 

\section{Comparison of Mean Euclidean Distance and Energy Score as Loss Functions}\label{appendix:mse_loss}
To compare the use of the mean Euclidean distance and the energy score as ABC loss functions, we run MCMC chains targeting each posterior. We use MCMC rather than SMC as it is challenging to fairly select stopping points to compare two different target posteriors. For the scoring rule posterior, we target,
\begin{equation} \label{eqn:SRPost}
p(\theta, x  \mid y) \propto \exp \left(- \omega \frac{1}{N}\sum_{n=1}^N \mathrm{SR}(y_i, x_{i, 1:M})\right) p(x|\theta) p(\theta).
\end{equation}
We set the learning rate $\omega=40$. This gives the loss function significant influence compared to the prior, without making the chains too sticky. When using the mean Euclidean distance, we replace $\mathrm{SR}(y_i, x_{i, 1:M})$ in Equation \ref{eqn:SRPost} with the Euclidean distance between $y_n$ and $x_n$, denoted $d_{\text{Euc}}$. Note that before calculating the Euclidean distance, as in the evaluation of the scoring rule, we first apply the function $\log(y+10)$ to the data and multiply by the weights assigned to each impact type due to measurement error (outlined in Section \ref{sec:measurement_error}). We keep the learning rate $\omega$ at 40 as the loss function here (mean Euclidean distance) converges to around 5, compared to the loss function in the SR-posterior which converges to around 4. Therefore both loss functions have similar weight compared to the prior in determining the posterior. 

\begin{figure}
\centering
\includegraphics[width=0.9\linewidth]{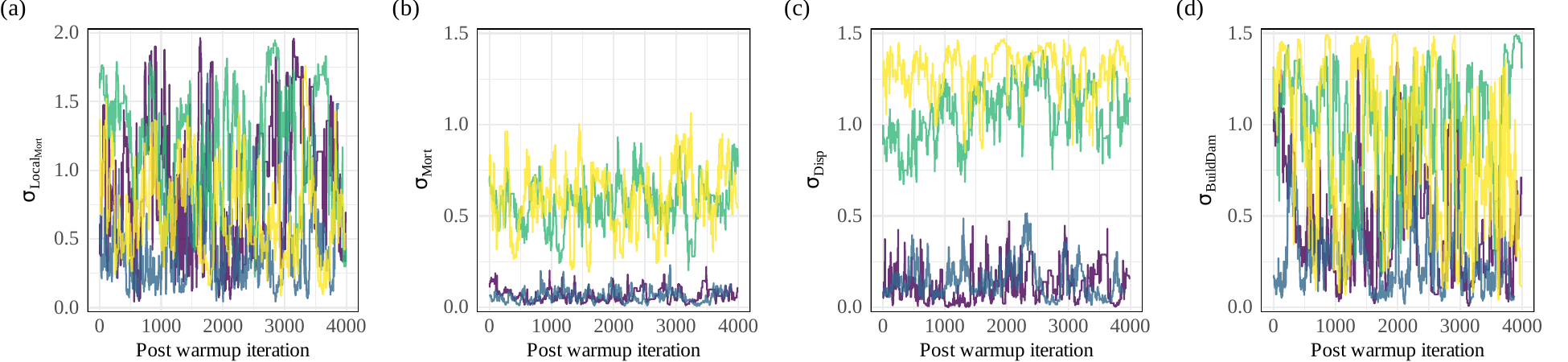}
\caption{Trace plots for the parameters determining the variance of the grid cell and event-wide error terms. The blue and purple trace plots use the Euclidean distance, and the yellow and green trace plots use the energy score. }
\label{fig:Trace_EuclDist}
\end{figure}

\begin{figure}
\centering
\includegraphics[width=0.75\linewidth]{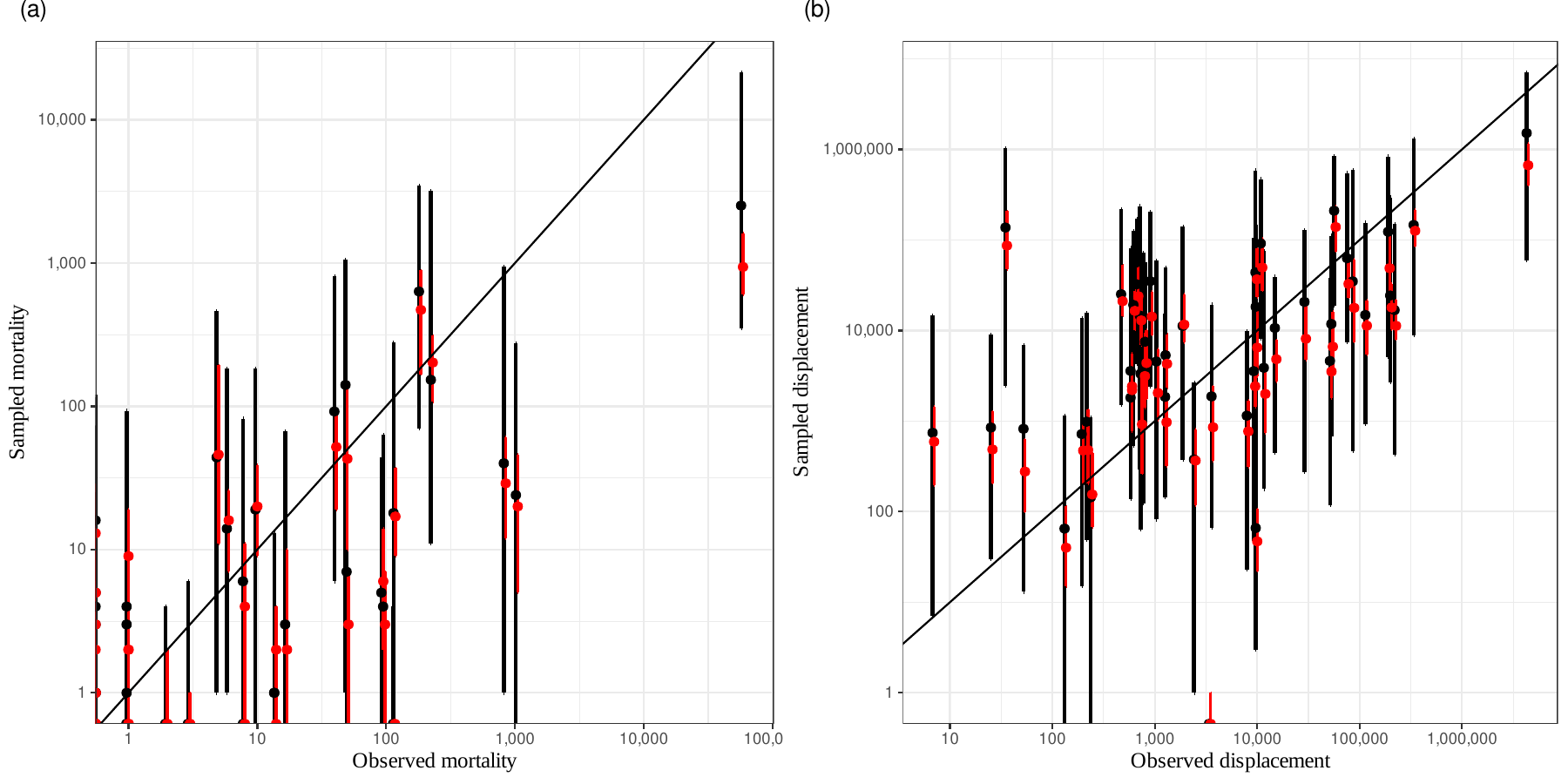}
\caption{Posterior predictive distributions for mortality and displacement under the models fitted using the energy score (black) and the Euclidean distance (red). The bars display $90\%$ credible intervals and the points the posterior median.}
\label{fig:EucDist_PostPred}
\end{figure}

We run the chains for $8000$ iterations, $4000$ of which we treat as the warmup period. We use the accelerated shaping and scaling adaptive MCMC algorithm proposed in \citet{spencer2021}. As described in \citet{dahlin2015}, we introduce a correlation of $0.9$ between samples of the event-wide error terms in the proposal and current iteration. We calculate the scoring rule using $M=100$ samples, and in calculating the Euclidean distance we average over 10 pseudo-marginal samples (note that this differs from $N$, which represents the number of events). In Figure \ref{fig:Trace_EuclDist} we compare trace plots for $\sigma_{\text{Local}_{\text{Mort}}}$, $\sigma_\text{Mort}$, $\sigma_\text{Disp}$, and $\sigma_\text{BuildDam}$ for 4 MCMC chains, two targeting the posterior in Equation \ref{eqn:SRPost} and two targeting the posterior using the Euclidean distance. It is evident that the Euclidean distance function results in underestimation of the variance terms in the event-wide error terms, particularly mortality and displacement (neither approach seems able to identify $\sigma_\text{BuildDam}$ due to the noisiness of the building damage data). In Figure \ref{fig:EucDist_PostPred} we also compare the posterior predictive distributions over the testing data for mortality and displacement. These plots further demonstrate that the Euclidean distance results in poorly calibrated posterior predictive distributions, with few of the $90\%$ credible intervals containing the observations. 

\section{Higher-level prior used with simulated data}\label{appendix:HLPrior_SimData}
When fitting the model on the simulated data, the vulnerability covariates are also included in the higher-level prior. To achieve this, we find the extremes of the full vulnerability component that can be attained using combinations of the 1\textsuperscript{st} and 99\textsuperscript{th} percentiles of the vulnerability covariates. We then calculate the impact probabilities at intensities $4.6$, $7$ and $9.5$ using these extremes, and ensure they fit within reasonable bounds. To accommodate the added variation introduced, the bounds on the impact probabilities are more relaxed, as detailed in Table \ref{table:hlp_bounds_simdata}. When working with the real data, it was found that considering all vulnerability covariates to be simultaneously at their extremes is not realistic and eliminated parameterisations with high posterior probability. This component of the higher-level prior was therefore removed when fitting the model on the real data, and we instead worked with the covariates at their standardised mean of zero but with less relaxed bounds. 

\begin{table}
\centering
\begin{tabular}{ r @{\hskip 15pt} r @{\hskip 15pt} r @{\hskip 15pt} r @{\hskip 15pt} r }
 \hline
 \rule{0pt}{3ex} 
 Intensity & $p^{\text{Mort}}$ & $p^{\text{Disp}} + p^{\text{Mort}}$ 
 & $p^{\text{BuildDam}}$ & $p^{\text{Disp}}$  \\
 $4.6$ & (0,0.03) & (0,0.1) & (0,0.15)  &  - \\ 
 $7$ & (0,0.15) & (0, 0.6) & ($10^{-6}$, 0.75)  & - \\ 
 $8$ & - & - & - & ($p^{\text{Mort}}$, 1)  \\ 
 $9.5$ & ($10^{-6}$,1) & (0.2,1) & (0.3,1) & -  \\
 \hline
\end{tabular}
 \caption{Bounds in the form (lower bound, upper bound) used in the higher level priors for the simulated data. Due to the relationship between the displacement and mortality probabilities, most checks are performed on their sum rather than the displacement probability itself, with an additional check that the probability of displacement is larger than the probability of mortality at an intensity of 8. }
 \label{table:hlp_bounds_simdata}
\end{table}

\section{Comparison of ABC-SMC and MCMC algorithms}\label{appendix:mcmc_comparison}
\begin{figure}
\centering
\includegraphics[width=0.95\linewidth]{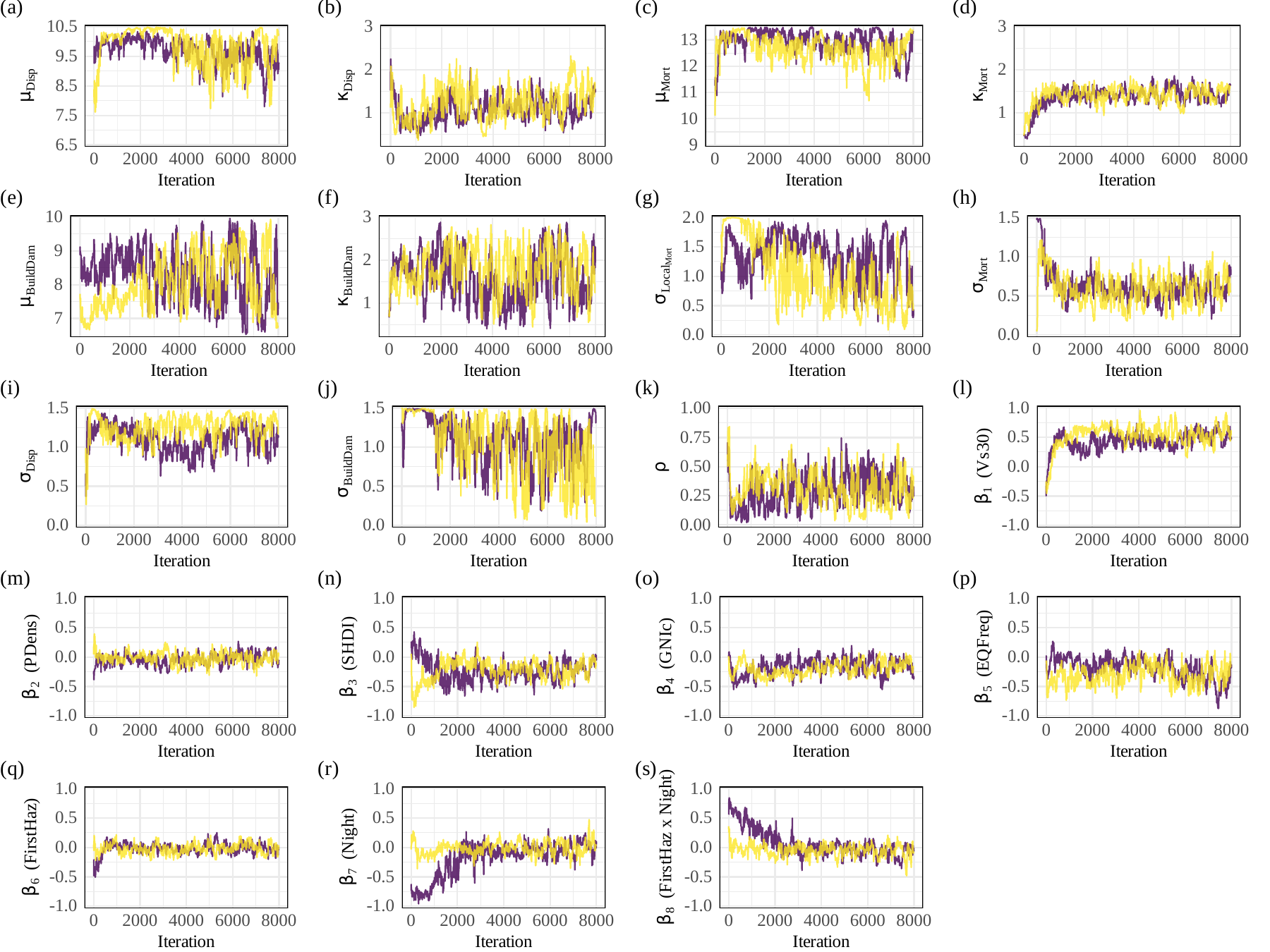}
\caption{Trace plot over 8000 iterations of two MCMC chains targeting the SR-posterior for the real data.}
\label{fig:Trace_MCMC}
\end{figure}

We run two MCMC chains for $8000$ iterations targeting the non-threshold-adjusted SR-posterior in Equation \ref{eqn:SRPost}. We use the accelerated shaping and scaling adaptive MCMC algorithm proposed in \citet{spencer2021}, and as described in \citet{dahlin2015}, we introduce a correlation of $0.9$ between samples of the event-wide error terms in the proposal and current iteration. As in Appendix \ref{appendix:mse_loss}, we set the learning rate $\omega_1 = 40$. Increasing the learning rate decreases the average acceptance rate, while decreasing the learning rate reduces the effect of the data compared to the priors. We find that $\omega_1 = 40$ provides a good balance between the two. We also chose this value to provide comparison to the ABC-SMC algorithm. The converged MCMC chain with $\omega_1 = 40$ results in evaluations of the loss function varying from around 3.6 to 3.8, while the final tolerance of the SMC algorithm is around 3.8. In Figure \ref{fig:Trace_MCMC} we display trace plots of two runs of the MCMC chains described. Both chains are started from parameterisations randomly drawn from the higher-level priors. While mixing is not perfect (we expect this is due partly to the correlation between the random error terms), there does appear to be agreement between the two chains.

\begin{figure}
\centering
\includegraphics[width=\linewidth]{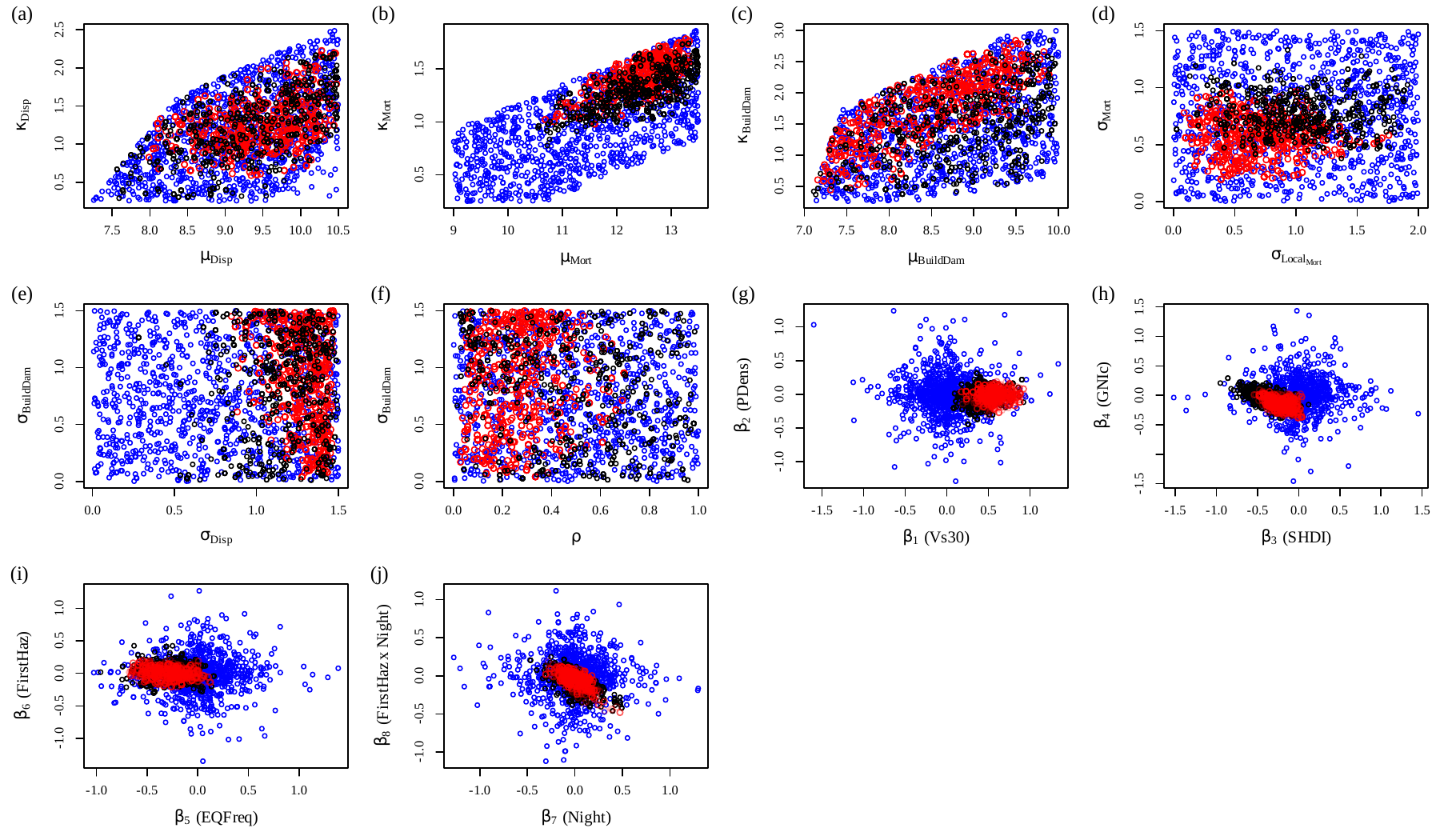}
\caption{Comparison of samples from the prior (shown in blue), with samples from the ABC-SMC posterior (shown in black) and the ABC-MCMC posterior (shown in red). The parameters are paired as many posteriors are correlated, but there also exists correlation beyond the pairings chosen.}
\label{fig:MCMCvsSMC}
\end{figure}

In Figure \ref{fig:MCMCvsSMC} we compare the posteriors obtained using ABC-MCMC to those from the ABC-SMC algorithm. To form the MCMC posterior, we take every 4th iteration after the warmup period of 4000 iterations. Note that the two approaches target two different posteriors: the ABC-MCMC results are for the standard SR-posterior outlined in \citet{pacchiardi2024}, while the ABC-SMC results are for a threshold-adjusted SR-posterior. Therefore, the ABC-SMC posterior is proportional to the prior when the loss function falls below the tolerance, while the posterior weight under the ABC-MCMC increases where the loss function decreases further. In general, we see a high level of agreement between the two approaches, particularly regarding the vulnerability covariates. The largest differences occur in the posteriors for the parameters $\sigma_{\text{Mort}}$ and $\rho$. The results for $\sigma_{\text{Mort}}$ could be attributed to the correlation in the event-wide error terms. By treating these random terms similarly to a latent variable, they can be optimised in a way that permits smaller variation. The results for $\rho$ are more surprising. The ability of the ABC-SMC algorithm to achieve similar losses with a greater range of values suggests that the ABC-MCMC algorithm may not be effectively exploring the parameter space of $\rho$. 

In most likelihood-free applications, including this one, sampling from the model is the most computationally demanding step in the algorithm. In the ABC-MCMC algorithm, the number of model samples is given by the number of chains multiplied by the number of iterations in each chain. If working with $3$ chains and $10,000$ iterations per chain, this gives $30,000$ samples from the likelihood. In the ABC-SMC algorithm, the number of samples is given by the number of steps multiplied by the number of particles. We further multiply by $\frac{3}{4}$ to roughly account for particles with zero weight at a given step. If working with $1000$ particles and $170$ steps (as in Section \ref{sec:Results}), this gives $127,500$ samples from the likelihood. Therefore, for our application, the ABC-SMC algorithm requires around $4$ times as many likelihood samples. However, the ABC-SMC algorithm can be parallelised between particles at each step, whereas the ABC-MCMC can only be parallelised between chains, with further shared-memory parallelisation within each model sample leading to less significant speedup. We have therefore found similar overall runtime between the two algorithms working with $50$ CPUs on a high-performance computing cluster. On a single CPU, the time to sample from the model once is around $12$ minutes, leading to an overall runtime of both algorithms of around 3 weeks.

\end{appendices}



\section{Acknowledgments}
The authors would like to thank the Internal Displacement Monitoring Center (IDMC), specifically Sylvain Ponserre, Maria-Teresa Miranda Espinosa, Bina Desai and Justin Ginnetti for the fruitful discussions with respect to population displacement data. This research was partially funded by the Engineering and Physical Sciences Research Council (EPSRC) Impact Acceleration Account Award EP/R511742/1. Max Anderson Loake acknowledges funding from the Rhodes Trust under the Rhodes Scholarship, and the EPSRC CDT in Modern Statistics and Statistical Machine Learning through grant EP/S023151/1.

For the purpose of Open Access, the authors have applied a CC BY public copyright licence to any Author Accepted Manuscript version arising from this submission.

This document uses the Oxford University Press LaTeX authoring template.

\end{document}